\documentclass[superscriptaddress, nofootinbib, amsmath, amssymb]{revtex4}
\usepackage{graphicx}
\usepackage{wasysym}
\usepackage[colorlinks=true,citecolor=blue,linkcolor=blue,urlcolor=blue]{hyperref}
\usepackage{color}
\usepackage[thinlines]{easytable}
\usepackage{multirow}
\usepackage{array}

\newcolumntype{M}[1]{>{\centering\arraybackslash}m{#1}}
\newcolumntype{N}{@{}m{0pt}@{}}

\begin{document}
\title{The 1.5 post-Newtonian radiative quadrupole moment in the context of a nonlocal field theory of gravity}
\author{Alain Dirkes}
\email{alain.dirkes@education.lu}
\affiliation{Lyc\'{e}e Ermesinde Mersch (LEM),\\ Rue de la Gare 45, Mersch, L-7590, Luxembourg\vspace{0.2cm}}
\vspace{0.9cm}

\affiliation{Frankfurt Institute for Advanced Studies (FIAS), Goethe University Frankfurt,\\ Ruth-Moufang Str.1, Frankfurt am Main, D-60438, Germany}

\begin{abstract}
We recently suggested a nonlocal modification of Einstein's field equations in which Newton's constant $G$ was promoted to a covariant differential operator $G_\Lambda(\Box_g)$. The latter contains two independent contributions, which operate respectively in the infrared (IR) and ultraviolet (UV) energy regimes. In the light of the recent direct gravitational radiation measurements we aim to determine the UV-modified 1.5 post-Newtonian radiative quadrupole moment of a generic $n$-body system. We eventually use these initial results in the particular context of a binary system and observe that in the limit vanishing UV parameters we precisely recover the corresponding general relativistic results. Moreover we observe that the leading order deviation of the UV-modified radiative quadrupole moment numerically coincides with findings obtained in the framework of calculations performed previously in the context of possible deviations of the perihelion precession of Mercury.

\end{abstract}
\maketitle
\section{Introduction}
\label{Introduction}
In chapter \ref{Introduction} we summarize the results outlined in \cite{Alain1} before we work out, in chapter \ref{UVQuadrupole}, the 1.5 post-Newtonian radiative quadrupole moment of a generic $n$-body system in the context of a particular nonlocal field theory of gravity. We close the discussion of this manuscript, in chapter \ref{Conclusion}, by a brief discussion of the of the main results obtained in this article.

\subsection{The nonlocally modified Einstein field equations}
The existence of a direct correlation between the gravitational field and a matter source term was discovered long before Albert Einstein published his famous field equations, $G_{\alpha\beta}\,=\,\frac{8\pi}{c^4} \ G \ T_{\alpha\beta}$, where $G_{\alpha\beta}$ is the Einstein curvature tensor and $T_{\alpha\beta}$ the energy-momentum tensor \cite{Einstein1}. Nevertheless it was only Einstein's theory of general relativity (GR) which was able, via the concept of spacetime curvature, to provide a deeper understanding of the true nature of gravity. Only one year after the final formulation of his theory of gravity, Einstein predicted the existence of gravitational waves, generated by time variations of the mass quadrupole moment of the source. Although the direct experimental detection is extremely challenging because of the waves' remarkably small amplitude \cite{Einstein2, Einstein3}, gravitational radiation has been measured indirectly since the mid seventies of the past century in the context of binary systems \cite{Taylor1, Burgay1, Stairs1, Stairs2, Taylor2}. Precisely one century after Einstein's theoretical prediction, an international collaboration of scientists (LIGO Scientific Collaboration and Virgo Collaboration) reported the first direct observation of gravitational waves \cite{LIGO1,LIGO2,LIGO3,LIGO4,LIGO5,LIGO6,LIGO7}. The first direct wave signal GW150914 was detected independently by the two LIGO detectors and its main features point to the coalescence of two stellar black holes. Three months later the same collaboration was able to perform a second direct gravitational wave measurement GW151226 allowing for an even better estimation of the stellar black hole population as well as for more robust constraints on possible general relativity deviations \cite{LIGO4}. Quite recently the two American detectors of the LIGO Scientific Collaboration as well as the European detector of the VIRGO Collaboration were able to simultaneously perform, on two different continents, direct gravitational waves measurements (GW170814) \cite{LIGO6}. This first simultaneous measurement was supplemented by the gravitational wave signal (GW170817), detected separately, both by the LIGO and the Virgo interferometers. This particular signal was the first observation of a binary neutron star inspiral \cite{LIGO7}. In this regard it is no surprise that this year's Nobel Prize in Physics was awarded to three of LIGO's most renowned pioneers, Barry C. Barish (Caltech), Kip S. Throne (Caltech) and Rainer Weiss (MIT)\footnote{Caltech: California Institute of Technology, MIT: Massachusetts Institute of Technology}. Despite the great success of Einstein's theory in describing the gravitational field some challenges remain yet unsolved. The most prominent questions are the dark energy and dark matter problems, the physical interpretation of black hole curvature singularities or the question of how to unify quantum mechanics and general relativity. In order to circumvent some of these issues many potentially viable alternative theories of gravity have been developed over the past decades \cite{Will1, Esposito1, Clifton1, Tsujikawa1, Woodard1, BertiBuonannoWill}.

\subsubsection{Nonlocal interactions}
The subject of field theories containing nonlocal interactions is a rather old idea that regained more and more attention over the past recent years. In this regard early attempts were made to introduce the concept of nonlocality into quantum field theory in order to remove ultraviolet divergencies occurring in the perturbation expansion of the S-matrix \cite{PeskinSchroeder,Wataghin,Efimov1,Efimov2,Namsrai,Tomboulis1,Tomboulis2,Tomboulis3,UVNonlocal}. Despite the success of perturbative renormalization in quantum electrodynamics (QED) in the late forties, the idea that local interactions may only be a low energy approximation to a more fundamental underlying theory of nonlocal interactions continued to be prominent in the fifties \cite{Tomboulis1, Tomboulis2,Tomboulis3,PaisUhlenbeck1, Yukawa1, Efimov1,Efimov2}. More recently the subject regained strong interest in connection with nonlocal theories of gravity, nonlocal models of cosmology as well as the nonlocality appearing in string field theory vertices \cite{Modesto5,Modesto8,NLSFT1,NLSFT2}. It was shown that nonlocal modifications can achieve ultraviolet finiteness or at least lead to superrenormalizability in the presence of gauge interactions. At the same time unitarity can be preserved, at least perturbatively, for thoroughly chosen analyticity conditions that have to be imposed on the nonlocal interactions. However, the causality issue remains a central concern, both in the classical as well as in the quantum theory. A way to physically motivate and to introduce the concept of nonlocality into field theories is to assume that spacetime itself has some kind of graininess at a very fundamentally small length scale \cite{Spallucci1,Namsrai,Snyder1,Maziashvili2,Maziashvili5}. In this context one can assume that the stochastic nature of spacetime could than manifest itself in the  interactions between fields and could be introduced, via nonlocal terms, at the level of the interaction Lagrangian of different fields. Let us illustrate this particular idea by an example borrowed from quantum field theory,
\begin{equation}
\label{NonlocalLagrangian}
\mathcal{L}(x)\,=\,\frac{1}{2}\varphi(x)\big(\Box-m^2\big)\varphi(x)-g\Big[\int d^4y\ K(x-y)\ \varphi(y)\Big]^4,
\end{equation} 
where $K(x-y)$ is the delocalization kernel (nonlocal form factor) defined, $K(x-y)=K(l^2 \Box)\delta^4(x-y)=\sum_{n=0}^{+\infty}\frac{c_n}{(2n)!}(l^2\Box)^n\delta^4(x-y)$, in terms of the generalized functions \cite{Namsrai,Tomboulis1, Tomboulis2,Tomboulis3,PaisUhlenbeck1, Yukawa1, Efimov1,Efimov2}. Here $l$ is a parameter of dimension length, $c_n$ are coefficients of the series expansion and $\Box$ is the flat spacetime d'Alembert operator. We observe from the interaction term in the Lagrangian, outlined in equation \eqref{NonlocalLagrangian}, that the field is not evaluated at a single point in spacetime, thus the interaction is nonlocal. As already mentioned previously, some authors believe that the idea of nonlocality is intimitatedly related to some kind of roughness of the spactime manifold itself. Especially in the context of theories of unified interactions the possibility was put forward that spacetime might be endowed with some sort of minimal length leading to a discretization of the manifold at a very fundamental scale \cite{Spallucci1,Namsrai,Snyder1,Maziashvili2,Maziashvili5}. As we have already claimed previously, the subject of nonlocal modifications in field theories has a rather long history and in this sense many different nonlocally modified theories are currently under investigation. Some of these theories originate from a top-down approach, in which theories that try to unify the fundamental interactions of physics exhibit a possible nonlocality, while others arise from purely bottom-up considerations \cite{Dvali2,Barvinsky1,Barvinsky2,Barvinsky3,Barvinsky4,Woodard1,Woodard2, Modesto5,Modesto6,Modesto7,Modesto8,Modesto9,HamberWilliams1,HamberWilliams2,Elizalde1,Elizalde2,Odintsov,Esposito1,Esposito2, Maggiore2,Maggiore3,Maggiore4}.

\subsubsection{The nonlocal field equations}
In this context we will, in the remaining part of this chapter, resume the main features of the nonlocally modified theory of gravity outlined for the first time in \cite{Alain1}. It should be observed that the main difference between our field equations and the standard theory of gravity is that in our approach Newton's gravitational constant is promoted to a covariant differential operator,
\begin{equation}
\label{NonlocalEinstein}
G_{\alpha\beta}\,=\, \frac{8\pi}{c^4} \ G_\Lambda(\Box_g) \ T_{\alpha\beta},
\end{equation}
where $\Box_g=\nabla^{\alpha}\nabla_\alpha$ is the covariant d'Alembert operator and $\sqrt{\Lambda}$ is the scale where infrared (IR) modifications become important. The covariant d'Alembert operator is sensitive to the characteristic wavelength of the gravitating system under consideration $1/\sqrt{-\Box_g} \sim \lambda_c$. We will see that our precise model will be constructed in such a way that the long-distance modification (IR term) is almost inessential for processes varying in spacetime much faster than $1/\sqrt{\Lambda}$ and large for slower phenomena at wavelengths $\sim \sqrt{\Lambda}$ and larger. In this regard spatially extended processes varying very slowly in time, with a small characteristic frequency $\nu_c\sim 1/\lambda_c$, will produce less spacetime curvature than smaller fast moving objects which couple to the gravitational field in almost the usual way. Cosmologically extended processes with a small characteristic frequency will effectively decouple from the gravitational field. John Wheeler's famous statement about the mutual influence of matter and spacetime curvature remains of course true, the precise form of the coupling differs however according to the dynamical nature of the gravitating object under consideration. Indeed promoting Newton's constant $G$ to a differential operator $G_{\Lambda}(\Box_g)$ allows for an interpolation between the Planckian value of the gravitational constant and its long distance magnitude \cite{Barvinsky1,Barvinsky2}, $G_P>G_\Lambda(\Box_g)>G_{L}$. Thus the differential operator acts like a high-pass filter with a macroscopic distance filter scale $\sqrt{\Lambda}$. In this way sources characterized by characteristic wavelengths much smaller than the filter scale  ($\lambda_c\ll\sqrt{\Lambda}$) pass undisturbed through the filter and gravitate normally, whereas sources characterized by wavelengths larger than the filter scale are effectively filtered out \cite{Dvali1,Dvali2}. In a more quantitative way we can see how this filter mechanism works by introducing the dimensionless parameter, $z\,=\,-\Lambda \Box_g\sim \Lambda/\lambda_c^2$,
\begin{equation}
G(z)\rightarrow G, \ |z|\gg 1 \ (\lambda_c \ll 1),\quad \quad G(z)\rightarrow 0, \ \  |z|\ll 1 \ (\lambda_c \gg 1).
\end{equation}
For small and fast moving objects with large values of $|z|$ (small characteristic wavelengths) the covariant coupling operator will essentially reduce to Newton's constant $G$, whereas for slowly varying processes characterized by small values of $|z|$ (large characteristic wavelengths) the coupling will be much smaller.  Despite the fact that the equations of motion $\eqref{NonlocalEinstein}$ are themselves generally covariant, they cannot, for nontrivial $G_\Lambda(\Box_g)$, be represented as a metric variational derivative of a diffeomorphism invariant action. The solution to this problem was suggested in \cite{Barvinsky1,Barvinsky2, Modesto1} by considering equation $\eqref{NonlocalEinstein}$ only as a first, linear in the curvature, approximation for the correct equations of motion. To specify our discussion we introduce at this stage the precise form of the covariant coupling operator used in this manuscript, 
\begin{equation}
G_{\Lambda}(\Box_g)\,=\,\mathcal{G}_{\kappa}(\Box_g) \cdot \mathcal{F}_\Lambda(\Box_g).
\end{equation}
We observe that $G_{\Lambda}(\Box_g)$ is composed by an ultraviolet (UV) modification term, $\mathcal{G}_{\kappa}=\frac{G}{1-\sigma e^{\kappa\Box_g}}$, and an infrared (IR) contribution, $\mathcal{F}_\Lambda=\frac{\Lambda \Box_g}{\Lambda \Box_g-1}$. It should be noticed that in the limit of vanishing wavelengths or infinitely large frequencies, we recover Einstein's theory of general relativity as the UV-term reduces to the Newtonian coupling constant ($\lim_{z\rightarrow +\infty} \mathcal{G}_{\kappa}(z)=G$) and the IR-term goes to one ($\lim_{z\rightarrow +\infty}\mathcal{F}_\Lambda=1$). The IR-degravitation essentially comes from $\lim_{z\rightarrow 0}\mathcal{F}_\Lambda(z)=0$ while the UV-term $\lim_{z\rightarrow 0}\mathcal{G}_\kappa(z)=\frac{G}{1-\sigma}$ taken alone does not vanish in this particular limit. In table \ref{TableStrengths} we outline the characteristic wavelengths of three physical systems operating in different energy regimes. For each of the three energy processes we present the deviations in the coupling strengths generated by the UV and IR covariant differential operators. It is interesting to observe that for microscopic objects, characterised by short wavelengths, the deviations caused by both operators are negligible. For astrophysical systems, like the Double Pulsar system \cite{Stairs1,Stairs2}, the effect caused by the UV term is 32 orders of magnitude stronger than the one generated by the IR operator. 
\begin{table}[h]
\begin{TAB}(r,1cm,1.3cm)[14pt]{|l|c|c|c|}{|c|c|c|c|c|}
Scale:     & Microscopic                                        & Astrophysical                            & Cosmological                           \\ 
System:    & Proton                                             &Double Pulsar                             &  Vacuum Energy                         \\ 
Wavelength:\hspace{0.2cm} & $\lambda_P\sim 10^{-15}\ m$         &$\lambda_D\sim 10^{+12}\ m$               & $\lambda_V\sim 10^{+30}\ m$            \\  
IR-term:   &\hspace{0.2cm} $\big|1-\mathcal{F}_\Lambda(\lambda_P)\big|\sim 10^{-90}\hspace{0.4cm}$&\hspace{0.2cm}$\big|1-\mathcal{F}_\Lambda(\lambda_D)\big|\sim 10^{-36}$\hspace{0.9cm}& $\big|1-\mathcal{F}_\Lambda(\lambda_V)\big|\sim 0.5$\\ 
UV-term:   &\hspace{-0.15cm} $\big|G-\mathcal{G}_\kappa(\lambda_P)\big| \sim 10^{-78} $\hspace{0.9cm}&$\big|G-\mathcal{G}_\kappa(\lambda_D)\big|\sim 10^{-4}$\hspace{0.9cm}& $\hspace{0.2cm}\big|G-\mathcal{G}_\kappa(\lambda_V)\big|\sim 10^{-4}$ \\ 
\end{TAB}
\caption{IR and UV deviations from the standard coupling (GR) between the gravitational field and three physical systems operating on different scales. The precise form of the IR and UV terms are, $\mathcal{F}_\Lambda(\lambda)=\frac{\Lambda/\lambda^2}{\Lambda/\lambda^2+1}$, $\mathcal{G}_\kappa(\lambda)=G\ [1-\sigma e^{-\kappa/\lambda^2}]^{-1}$, respectively and $\Lambda=10^{60}\ m^2$, $\sigma=2\cdot 10^{-4}$ and $\kappa=5\cdot 10^{-3}\ m^2$.}\label{TableStrengths}
\end{table}
However in the context of the vacuum energy the situation is completely reversed. While the UV deviation remains essentially the same, the IR term leads to a strong modification in the coupling strength when compared to standard theory of gravity (GR). Moreover it should be noticed that, in contrary to most of the UV terms studied in the literature, our term is most sensible to intermediate wavelengths or larger. In this regard we do not observe a UV {\it degravitation} in the limit of infinitely rapid energy processes but we merely recover in this limit the standard field equations, $\lim_{+\infty}G_\kappa(z)=G$. On the other side this term is not an IR term either as we do not observe an IR {\it degravitation}, $\lim_{z\rightarrow 0}\mathcal{G}_\kappa(z)=0$, in the limit of slowly varying energy processes. 
The UV term discussed in this manuscript actually interpolates between a strong UV (SUV) term, where we observe an UV {\it degravitation}, $\lim_{z\rightarrow +\infty} G_{SUV}(z)=0$, and an IR term characterised by an IR {\it degravitation}, $\lim_{z\rightarrow 0} \mathcal{F}_\Lambda(z)=0$. A more detailed discussion about the simplest SUV term ($\lim_{z\rightarrow +\infty}e^{-\frac{l^2_{UV}}{\Lambda}z}=0$) can be found in \cite{Modesto1, Modesto2, Modesto3}. From an astrophysical point of view it should however be noticed that the deviation in the coupling strength caused by this SUV term, $|G-G\ e^{-l^2_{UV}/\lambda^2_D}|\approx0$, is rather small compared to deviations (Table \ref{TableStrengths}) generated by our soft UV operator. For completeness we also remind that in the limit of a small UV parameter $\kappa$ and an infinitely extended Universe, $\frac{\Lambda}{\kappa}\rightarrow \infty$, the (soft) UV poles are removed from the complex plane and unitarity (no ghosts) is restored \cite{Alain1}. Finally it should be remarked that the covariant d'Alembert operator $\sqrt{-\Box_g}\sim 1/\lambda_c$ does not only {\it measure} the spacetime variations of energy processes but is also sensitive to the amount of spacetime curvature produced by the object under consideration. Therefore the results presented in table \ref{TableStrengths} are only approximately correct in the sense that in addition to the scale discussion outlined above we have to take into account that nonlocal modifications become more important with increasing spacetime curvature.

\subsubsection{The nonlocal wave equation}
\label{EWE-16.05.17}
We saw in \cite{Alain0A,Alain0B,Alain1} that the nonlocally modified wave equation naturally originates from the quest of sharing out some of the complexity of the nonlocal coupling operator $G(\Box_g)$ to both sides of the relaxed Einstein equations. We have shown previously in this subsection that it is possible to split the nonlocal coupling operator, acting on the matter source term $T^{\alpha\beta}$, into a flat space contribution $G(\Box)$ multiplied by a highly nonlinear differential piece $\mathcal{H}(\Box,w)$. We aim to summarize first what this means for the effective energy-momentum tensor, $\mathcal{T}^{\alpha\beta}=G(\Box) \ \mathcal{H}(w,\partial) \ T^{\alpha\beta}$. In the pursuit of removing some of the differential complexity from the effective energy-momentum tensor $\mathcal{T}^{\alpha\beta}$ we will apply the inverse flat spacetime operator $G^{-1}(\Box)$ to both sides of the relaxed Einstein field equation, $G^{-1}(\Box) \ \Box h^{\alpha\beta}\,=\,-\frac{16 \pi G}{c^4} \ G^{-1}(\Box) \big[(-g)\mathcal{T}^{\alpha\beta}+\tau_{LL}^{\alpha\beta}+\tau_H^{\alpha\beta}\big]$. We will see that it is precisely this mathematical operation which will finally lead us to derive the nonlocal wave equation,
\begin{eqnarray}
\Box_{c} \ h^{\alpha\beta}(x)\,=\, -\frac{16 \pi G}{c^4}N^{\alpha\beta}(x).
\end{eqnarray}
where $\Box_{c}$ is the effective d'Alembert operator $\Box_{c}=\big[1-\sigma e^{\kappa\Delta}\big] \ \Box$ \cite{Alain0A,Alain0B,Alain1,Alain2}. $N^{\alpha\beta}$ is a pseudotensorial quantity  which we will call in the remaining part of this article the effective energy-momentum pseudotensor, $N^{\alpha\beta}=G^{-1}(\Box) \big[(-g)\mathcal{T}^{\alpha\beta}+\tau_{LL}^{\alpha\beta}+\tilde{\tau}_H^{\alpha\beta}\big]$, where $\tilde{\tau}^{\alpha\beta}_m=(-g)\mathcal{T}^{\alpha\beta}$ is the effective matter pseudotensor, $\tau_{LL}^{\alpha\beta}=(-g)t_{LL}^{\alpha\beta}$ is the Landau-Lifshitz pseudotensor and $\tilde{\tau}_H^{\alpha\beta}=(-g)t^{\alpha\beta}_H+G(\Box)\mathcal{O}^{\alpha\beta}(h)$ is the effective harmonic gauge pseudotensor where $\mathcal{O}^{\alpha\beta}(h)=-\sigma\sum_{n=1}^{+\infty}\frac{(\kappa)^n}{n!}\partial^{2n}_0 e^{\kappa\Delta} \Box h^{\alpha\beta}$ is the iterative post-Newtonian potential correction contribution. This term is added to the right-hand-side of the wave equation very much like the harmonic gauge contribution is added to the right-hand-side for the standard relaxed Einstein equation \cite{PoissonWill, PatiWill1, PatiWill2, Blanchet1, Will1}. It should be noticed that the modified d'Alembert operator $\Box_{c}$ is of the same post-Newtonian order than the standard d'Alembert operator, $\Box_c=\mathcal{O}(c^{-2})$ and reduces to the usual one in the limit of vanishing UV modification parameters, $\lim_{\sigma,\kappa \rightarrow 0}\ \Box_{c}\,=\, \Box$. In the same limits the effective pseudotensor $N^{\alpha\beta}$ reduces to the general relativistic one, $\lim_{\sigma,\kappa \rightarrow 0}  \ N^{\alpha\beta}=\tau^{\alpha\beta}$. The second limit is less straight forward, but from the precise form of $\mathcal{T}^{\alpha\beta}$ as well as from the inverse differential operator $G^{-1}(\Box)$ we can see that we recover the usual effective energy-momentum pseudotensor, $\tau^{\alpha\beta}=\tau^{\alpha\beta}_m+\tau^{\alpha\beta}_{LL}+\tau^{\alpha\beta}_H$. Further conceptual and computational details on this very important quantities are provided in \cite{Alain1}. At the level of the wave equations, these two properties can be summarized by the following relation,
\begin{equation}
\Box_c \ h^{\alpha\beta}(x)\,=\, -\frac{16 \pi G}{c^4}N^{\alpha\beta}(x) \ \ \underset{\sigma,\kappa\rightarrow 0}{\Longrightarrow} \ \ \Box \ h^{\alpha\beta}(x)\,=\, -\frac{16 \pi G}{c^4}\tau^{\alpha\beta}(x).
\end{equation}
In order to solve this equation we will use, in analogy to the standard wave equation, the following ansatz, $h^{\alpha\beta}(x)\,=\,-\frac{16 \pi G_N}{c^4} \int d^4y \ G(x-y) \ N^{\alpha\beta}(y)$, together with the identity for the effective Green function, $\Box_{c} G(x-y)\,=\,\delta(x-y)$, to solve for the potentials $h^{\alpha\beta}$ of the modified wave equation. Following the usual procedure \cite{PoissonWill,Maggiore1,Buonanno1}, presented in \cite{Alain1}, we obtain the Green function in momentum space,
\begin{equation}
\label{GreenMomentum}
G(k)\,=\, \frac{1}{(k^0)^2-|\textbf{k}|^2}+\sigma \ \frac{ \ e^{-\kappa|\textbf{k}|^2}}{(k^0)^2-|\textbf{k}|^2}+\cdots.
\end{equation}
It should be noticed that the first of these two contributions will eventually give rise to the usual Green function \cite{Alain0A,Alain0B,Alain1,Alain2}. These considerations finally permit us to work out an expression for the retarded Green function, $G_r(x-y)\,=\,G_r^{GR}+G_r^{NL}$, where, $G_r^{GR}=\frac{-1}{4\pi}\frac{\delta(x^0-|\textbf{x}-\textbf{y}|-y^0)}{|\textbf{x}-\textbf{y}|}$, is the well known retarded Green function and $G_r^{NL}=\frac{-1}{4\pi}\frac{1}{|\textbf{x}-\textbf{y}|}\frac{\sigma}{2\sqrt{\kappa\pi}}e^{-\frac{(x^0-|\textbf{x}-\textbf{y}|-y^0)^2}{4\kappa}}$ is the nonlocal correction term. In this way we are able to recover in the limit of vanishing modification parameters the usual retarded Green function, $\lim_{\sigma,\kappa \rightarrow 0} \ G_r(x-y)=G_r^{GR}$. In addition it should be pointed out that we have, by virtue of the exponential representation of the Dirac distribution, the following result, $\lim_{\kappa\rightarrow 0} \frac{1}{2\sqrt{\kappa\pi}}e^{-\frac{(x^0-|\textbf{x}-\textbf{y}|-y^0)}{4\kappa}}\,=\, \delta(x^0-|\textbf{x}-\textbf{y}|-y^0)$. In analogy to the purely general relativistic case we can write down the formal solution to the modfied wave equation,
\begin{equation}
\label{ModSol-22.05.17}
h^{\alpha\beta}(x)\,=\,  \frac{4 \ G}{c^4}  \int d\textbf{y} \ \frac{N^{\alpha\beta}(x^0-|\textbf{x}-\textbf{y}|,\textbf{y})}{|\textbf{x}-\textbf{y}|}.
\end{equation}
The retarded effective pseudotensor can be decomposed into two independent pieces according to the two contributions coming from the retarded Green function, $N^{\alpha\beta}(x^0-|\textbf{x}-\textbf{y}|,\textbf{y})=\mathcal{D} N^{\alpha\beta}(y^0,\textbf{y})+\sigma \mathcal{E} N^{\alpha\beta}(y^0,\textbf{y})$, where for later convenience we introduced the following two integral operators, $\mathcal{D}= \int dy^0 \ \delta(x^0-|\textbf{x}-\textbf{y}|-y^0)$ and $\mathcal{E}=\int dy^0 \ \frac{1}{2\sqrt{\pi \kappa}} e^{-\frac{(x^o-|\textbf{x}-\textbf{y}|-y^o)^2}{4\kappa}}$.

\subsubsection{The effective nonlocal pseudotensor}
From \cite{Alain0A,AlainB,Alain1} we recall the precise expression for the matter contribution of the effective pseudotensor, $N_{m}^{\alpha\beta}=G^{-1}(\Box) \big[(-g) \ \mathcal{T}^{\alpha\beta}\big]=\big[G(\Box)\big]^{-1} \big[(-g) \ G(\Box) \ \mathcal{B}^{\alpha\beta}\big]$. In order to extract from this expression all the relevant pieces that lie within the order of accuracy that we aim to work at in this article, we essentially need to address two different tasks. In a first step we have to review the leading terms of $\mathcal{B}^{\alpha\beta}$ (chapter \ref{modifiedrelaxedEinsteinequations}) and see in how far they may contribute to the 1.5 post-Newtonian order of accuracy. In a second step we have to analyze how the differential operator $G^{-1}(\Box)$ acts on the product of the metric determinant $(-g)$ multiplied by the energy-momentum tensor $\mathcal{T}^{\alpha\beta}=G(\Box) \ \mathcal{B}^{\alpha\beta}$. Although this formal operation will lead to additional terms, the annihilation of the operator $G(\Box)$ with its inverse counterpart will substantially simplify the differential structure of the original energy-momentum tensor $\mathcal{T}^{\alpha\beta}$. Before we come to the two tasks mentioned above we first need to set in place a couple of preliminary results. From a technical point of view we need to introduce the operators of instantaneous potentials, $ \Box^{-1}[ \bar{\tau}]=\sum_{k=0}^{+\infty} \Big(\frac{\partial}{c\partial t}\Big)^{2k} \ \Delta^{-k-1}[\bar{\tau}]$ \cite{Blanchet1, Blanchet3, Blanchet4}. This operator is instantaneous in the sense that it does not involve any integration over time. However one should be aware that unlike the inverse retarded d'Alembert operator, this instantaneous operator will be defined only when acting on a post-Newtonian series $\bar{\tau}$. Another important computational tool which we borrow from \cite{Blanchet1, Blanchet3, Blanchet4} are the generalized iterated Poisson integrals, $\Delta^{-k-1}[\bar{\tau}_m](\textbf{x},t)=-\frac{1}{4\pi} \int d\textbf{y} \ \frac{|\textbf{x}-\textbf{y}|^{2k-1}}{(2k)!} \ \bar{\tau}_m(\textbf{y},t)$, where $\bar{\tau}_m$ is the $m$-th post-Newtonian coefficient of the energy-momentum source term, $\bar{\tau}=\sum_{m=-2}^{+\infty} \bar{\tau}_m/c^{m}$. An additional important result that needs to be mentioned is the generalized regularization prescription\footnote{The author would like to thank Professor E. Poisson for useful comments regarding this particular issue.}, $\big[\nabla^m \frac{1}{|\textbf{x}-\textbf{r}_A|}\big] \ \big[\nabla^n \delta(\textbf{x}-\textbf{r}_A)\big]\equiv 0, \ \forall n,m\in \mathbb{N}$. The need for this kind of regularization prescription merely comes from the fact that inside a post-Newtonian expansion, the nonlocality of the modified Einstein equations will lead to additional derivatives which act on the Newtonian potentials. It is easy to see that in the limit $m=0$ and $n=0$ we recover the well known standard regularization prescription \cite{PoissonWill, Blanchet1, Blanchet2}. We are now ready to come to the first of the two tasks mentioned in the beginning of this subsection. In order to extract the pertinent pieces from $\mathcal{B}^{\alpha\beta}= \mathcal{H}(w,\Box)\ \big[\tau_m^{\alpha\beta}/(-g)\big]$ to the required order of precision, we need first to have a closer look at the differential curvature operator $\mathcal{H}(w,\Box)$. We know \cite{Alain0A,Alain0B,Alain1} that it is essentially composed by the potential operator function $w(h,\partial)$ and the flat spacetime d'Alembert operator,
\begin{equation}
w(h,\partial)\,=\,-h^{\mu\nu} \partial_{\mu\nu}+\tilde{w}(h)\Box-\tilde{w}(h) h^{\mu\nu}\partial_{\mu\nu}\,=\,-\frac{h^{00}}{2}\Delta+\mathcal{O}(c^{-4}),
\end{equation}
where $\partial_{\mu\nu}=\partial_\mu\partial_\nu$. We see that at the 1.5 post-Newtonian order of accuracy, the potential operator function $w(h,\partial)$ reduces to one single contribution, composed by the potential $h^{00}=\mathcal{O}(c^{-2})$ \cite{WillWiseman, PatiWill1,PoissonWill,WagonerWill1976} and the flat spacetime Laplace operator $\Delta$. With this in mind we can finally take up the leading four contributions of the curvature energy-momentum tensor $\mathcal{B}^{\alpha\beta}$,
\begin{equation}
\begin{split}
\mathcal{B}^{\alpha\beta}_{1}\,&=\,\tau^{\alpha\beta}_{m}(c^{-3})-\tau^{\alpha\beta}_m(c^0) \ h^{00}+\mathcal{O}(c^{-4}),\\
B^{\alpha\beta}_{2}\,&=\,-\frac{\epsilon}{2}\sum_A m_A v^\alpha_A v^\beta_A \ \Big[\sum_{n=0}^\infty \sigma^n e^{(n+1)\kappa \Delta} \Big] \ \Big[h^{00}\Delta \delta(\textbf{y}-\textbf{x}_A)\Big]+\mathcal{O}(c^{-4}),\\
\mathcal{B}^{\alpha\beta}_3\,&=\,\frac{\epsilon\kappa}{2} e^{\kappa\Box}\Big[\frac{w^2}{1-\sigma e^{\kappa \Box}}\Big] \Big[\frac{\tau^{\alpha\beta}_m}{(-g)}\Big]\,\propto\, w^2\,=\mathcal{O}(c^{-4}),\\
\mathcal{B}^{\alpha\beta}_4\,&=\,\frac{\epsilon\kappa^2}{3!} e^{\kappa\Box}\Big[\frac{w^3}{1-\sigma e^{\kappa \Box}}\Big] \Big[\frac{\tau^{\alpha\beta}_m}{(-g)}\Big]\,\propto\, w^3\,=\mathcal{O}(c^{-6}).
\end{split}
\end{equation}
The terms $\mathcal{B}^{\alpha\beta}_3$ and $\mathcal{B}^{\alpha\beta}_4$ are beyond the order of accuracy at which we aim to work at in this article because $\omega^2=\mathcal{O}(c^{-4})$ as well as $\omega^3=\mathcal{O}(c^{-6})$ and $\tau_m(c^0)$ is the matter pseudotensor at the leading order of accuracy.  The time-time component of the nonlocal Landau-Lifshitz pseudotensor $N^{00}_{LL}=G^{-1}(\Box) \ \tau^{00}_{LL}$, where $\tau^{00}_{LL}=\frac{-7}{8\pi G} \partial_jV\partial^jV+\mathcal{O}(c^{-2})$ \cite{ WillWiseman, PatiWill1,PoissonWill}. We will see in the next chapter that this term will suffice to work out the physical quantities that we are interested in this thesis, 
\begin{eqnarray}
\label{champ-02.06.17}
c^{-2}N^{00}_{LL}\,=\, c^{-2}\Big[ \big(1-\sigma\big)\tau^{00}_{LL} -\epsilon \Delta \tau^{00}_{LL}-\sigma\sum_{m=2}\frac{\kappa^m}{m!}\Delta^m \tau^{00}_{LL}\Big]+\mathcal{O}(c^{-4}).
\end{eqnarray}
This result was derived by using a series expansion of the exponential differential operator and by taking into account that $\partial_0=\mathcal{O}(c^{-1})$. The effective Landau-Lifshitz tensor contribution was scaled by the factor $c^{-2}$ for later convenience. From the leading term we will eventually be able to recover the standard post-Newtonian field contribution \cite{Alain0A,Alain0B,Alain1}.

\subsection{The effective UV radiative quadrupole moment}
\label{UVQuadrupole}
We saw in Table \ref{TableStrengths} as well as in \cite{Alain1} that the IR-contribution of $G_\Lambda(\Box_g)$ does not lead to noticable deviations from general relativity on astrophysical scales. We therefore content ourselves in this subsection to work out only the UV modified near zone radiative quadrupole moment at the 1.5 post-Newtonian order of accuracy. In a first step the computations are performed for a system composed by isolated spinless bodies with masses $m_A$ ($A\in \mathbb{N}$). In a second step we rephrase these results in terms of the characteristic notation used in the context of binary systems. We know from \cite{Alain1} that the radiative quadrupole moment can be decomposed into a matter $Q_m^{ab}$ and a field contribution $Q^{ab}_{LL}$. In addition it was seen previously that the radiative quadrupole moment is a function of the retarded time,
\begin{equation*}
Q^{ab}\,=\,\Big(\mathcal{D}-\sigma\mathcal{E}\Big)\ c^{-2} \int_{\mathcal{M}}d\textbf{x} \ \Big(N^{00}_m+N^{00}_{LL}\Big) \ x^ax^b\,=\,\Big(\mathcal{D}-\sigma\mathcal{E}\Big)\ \Big(Q^{ab}_m+Q^{ab}_{LL}\Big).
\end{equation*} 
We also remind that the retarded time integral operators $\mathcal{D}$ and $\mathcal{E}$ originate from the modified Green function. Our strategy will be to review the matter and field pieces separately before we finally assemble the two independent contributions to what will be the UV modified near zone radiative quadrupole moment at the 1.5 post-Newtonian order of accuracy.
\subsubsection{The matter contribution: $Q^{ab}_m$}
We aim to review the different matter components one after the other and we will see in how far they will eventually contribute to the 1.5 post-Newtonian result. The first contribution that needs to be carefully looked at is the following one,
\begin{equation*}
\begin{split}
Q^{ab}_{\mathcal{B}_{1}+\mathcal{B}h}\,=&\,c^{-2}\int_{\mathcal{M}}d\textbf{x} \ \Big[\mathcal{B}^{00}_{1}+\mathcal{B}^{00}h^{00}\Big]\ x^ax^b\,=\,Q^{ab}_m[GR]+3\sigma \frac{G}{c^2} \sum_A\sum_{B\neq A} \frac{m_Am_B}{|\textbf{x}_A-\textbf{x}_B|} \ x^a_A x^b_A+\mathcal{O}(c^{-4}). 
\end{split}
\end{equation*}
It is straightforward to see that in the limit of vanishing modification parameters, we precisely recover the usual general relativistic contribution \cite{WagonerWill1976,WillWiseman,PatiWill1,PatiWill2,PoissonWill,LandauLifshitz},
\begin{equation*}
\lim_{\sigma\rightarrow 0} Q^{ab}_{\mathcal{B}_1+\mathcal{B}h}\,=\,Q^{ab}_m[GR]\,=\, \sum_A m_A \bigg[1+\frac{v^2_A}{2c^2}+3 \frac{G}{c^2}\sum_{B\neq A} \frac{m_B}{|\textbf{x}_A-\textbf{x}_B|}\bigg]x^a_A x^b_A+\mathcal{O}(c^{-4}).
\end{equation*}
From the previous subsection we pick up the precise expression for $\mathcal{B}^{00}_{2}$ and use it inside the quadrupole integral,
\begin{equation}
\begin{split}
Q^{ab}_{\mathcal{B}_{2}}\,=&\,c^{-2} \int_{\mathcal{M}}d\textbf{x} \ B^{00}_{A5} \ x^ax^b\,
=\,4\epsilon\frac{1+\sigma}{1-\sigma} \frac{G}{c^2} \sum_A\sum_{B\neq A} m_Am_B \ \bigg[\frac{r_{AB}^b x_A^{a}}{r^3_{AB}}  +\frac{r_{AB}^a x_A^b}{r_{AB}^3} -\frac{\delta^{ab}}{r_{AB}}\bigg]+\mathcal{O}(c^{-4}).
\end{split}
\end{equation}
It should be noticed that although this contribution contains a sum of infinitely many derivatives, the result at 1.5 post-Newtonian order of accuracy is finite. This is due to the fact that only the lowest order derivative terms contribute to the final result. Higher order derivative terms will lead, after partial integration(s), to contributions that are proportional to either one of the following two relations or to both relations simultaneously,
\begin{equation}
\begin{split}
\sum_A\sum_{B\neq A}m_A \tilde{m}_B \ \nabla^q\delta(\textbf{x}_A-\textbf{x}_B)\,=\,0,\quad \forall q\geq 0,\quad\nabla^q [x^ax^b]\,=\,0, \quad \forall q\geq 3.
\end{split}
\end{equation}
It should be noticed that surface terms, arising from partial integration, can be freely discarded in the near zone domain \cite{WillWiseman,PatiWill1,PatiWill2,PoissonWill}, $\int_{\partial \mathcal{M}} dS^p \ \partial_p h^{00}\frac{\delta(\mathbf{y}-\mathbf{x}_A)}{|\mathbf{x}-\mathbf{y}|}  \propto \delta(\mathcal{R}-|\mathbf{x}_A|)$. In addition we observe that with the use of the parameter $\epsilon=\sigma \kappa$ (dimension length-squared) the quadrupole moment has the right physical dimensions of mass times length-squared. Similar remarks apply to the derivative contribution $D^{\alpha\beta}$ as it involves, here again, infinitely many derivative terms \cite{Alain0A,Alain1}. This term originates from the computational process of the nonlocal annihilation of $G^{-1}(\Box)$ with $G(\Box)$ inside the effective nonlocal energy-momentum pseudotensor \cite{Alain1}. The result is however finite at the order of precision that we aim to work at in this thesis. Additional computational steps for this as well as for the previous result can be found in the appendix \ref{AppendixUVQuadrupole} related to this subsection,
\begin{equation}
\begin{split}
Q^{ab}_{D}\,=\,c^{-2}\int_{\mathcal{M}}d\textbf{x} \ D^{00} \ x^ax^b\,=\,-8\epsilon\frac{1+\sigma}{1-\sigma} \frac{G}{c^2} \ \sum_A \sum_{B\neq A}m_A  m_B \ \bigg[ \frac{x_A^a r_{AB}^b}{r_{AB}^3} +\ \frac{r_{AB}^a x_A^b}{r_{AB}^3}\bigg].
\end{split}
\end{equation}
We have seen previously that the additional terms are beyond the 1.5 PN order of precision. This allows us to write down the complete near zone matter radiative quadrupole moment in terms of the general relativistic component and the nonlocal contribution,
\begin{equation}
\begin{split}
Q^{ab}_m\,=&\,Q^{ab}_{m}[GR]+Q^{ab}_m[NL]+\mathcal{O}(c^{-4}).
\end{split}
\end{equation}
The precise expressions for $Q^{ab}_m[GR]$ and $Q^{ab}_{m}[NL]$, in the context of a general $n$-body system, can be found in the appendix \ref{AppendixUVQuadrupole} related to this subsection. For clarity reasons we prefer to present here only the results for a two-body system,
\begin{gather}
Q^{ab}_m[GR]\,=\,\eta m \Big[1+\frac{1}{2}(1-3\eta)\frac{v^2}{c^2}+3(1-2\eta)\frac{Gm}{c^2r}\Big],\\
Q^{ab}_{m}[NL]\,=\,\frac{\eta m}{c^2} \Big[3\sigma (1-2\eta)\frac{Gm}{r} -8g(\epsilon,\kappa)\Big(\frac{Gm}{r^3}r^ar^b+\frac{Gm}{r}\delta^{ab}\Big)\Big],
\end{gather}
where $m=m_1+m_2$ is the sum of the masses of the two isolated bodies, $\eta=(m_1m_2)/(m_1+m_2)^2$ is a dimensionless parameter, $r=|\textbf{x}_1-\textbf{x}_2|$ is the relative separation of the two bodies and $g(\sigma,\epsilon)=\epsilon (1+\sigma)/(1-\sigma)$ is a parameter function of dimension length-squared. Now that we have worked out the matter components to the required order of precision we will turn our attention to the field (Landau-Lifshitz) contributions.
\subsubsection{The field contribution: $Q^{ab}_{LL}$}
In the quest of working out the near zone field contribution of the radiative quadrupole moment $Q^{ab}_{LL}$ of generic $n$-body system, we need to employ the nonlocally moified Landau-Lifshitz pseudotensor $N^{\alpha\beta}_{LL}$ outlined previously,
\begin{equation}
\begin{split}
Q^{ab}_{LL}\,=&\,-\frac{7}{8\pi c^2G} \int_{\mathcal{M}} d\textbf{x} \ \Big[1-\sigma e^{\kappa\Delta}\Big] \ \Big[\partial_pV\partial^pV\Big] \ x^ax^b+\mathcal{O}(c^{-4}),
\end{split}
\end{equation}
where we remind from subsection that $V=(1+\sigma) \ U$ is the effective Newtonian potential and $U=\sum_A\frac{m_A G}{|\textbf{x}-\textbf{x}_A|}$ is the standard Newtonian potential. We would also like to recall that additional computational details for this and the remaining computations of this subsection can be withdrawn from the appendix \ref{AppendixUVQuadrupole} related to this subsection. The first term is proportional to the usual general relativistic 1.5 post-Newtonian near zone field contribution $Q^{ab}_{LL}[GR]$ \cite{WillWiseman, PatiWill1,PoissonWill},
\begin{equation}
\begin{split}
-\frac{7}{8\pi c^2G} \int_{\mathcal{M}} d\textbf{x} \ \partial_pV\partial^pV \ x^ax^b\,=&\,-\frac{7G}{2c^2}\sum_A\sum_{B\neq A} \frac{\tilde{m}_A\tilde{m}_B}{r_{AB}}x^a_Ax^b_A\,=\,(1+\sigma)^2 Q_{LL}[GR].
\end{split}
\end{equation}
The second term entails, through the exponential differential operator, a sum of infinitely many derivatives \cite{Alain2,Alain1},
\begin{equation}
\begin{split}
-\sigma\int_{\mathcal{M}} d\textbf{x} \ \Big[ e^{\kappa\Delta} \ \partial_pV\partial^pV\Big] \ x^a x^b\,=&\,-\sigma\int_{\mathcal{M}} d\textbf{x} \ \Big[ (1+\kappa \Delta+\frac{\kappa^2}{2}\Delta^2) \ \partial_pV\partial^pV\Big] \ x^a x^b.
\end{split}
\end{equation}
However a careful analysis (appendix \ref{AppendixUVQuadrupole}) shows that higher order derivative terms will lead, after partial integration(s), to contributions that are proportional to either one of the following two relations or to both relations at the same time,
\begin{equation}
\begin{split}
\sum_A\sum_{B\neq A}m_A \tilde{m}_B \ \nabla^q\delta(\textbf{x}_A-\textbf{x}_B)\,=\,0,\quad \forall q\geq 0,\quad\nabla^q [x^ax^b]\,=&\,0, \quad \forall q\geq 3.
\end{split}
\end{equation}
We remind that surface terms arising from partial integration can be freely discarded in the near zone. The remaining three contributions of this second term have to be analyzed separately. A closer look reveals that the first one is very similar to the one which has already been worked out in this subsection and leads to a contribution which is directly proportional to the 1.5 post-Newtonian general relativistic field contribution,
\begin{equation}
-\sigma \int_{\mathcal{M}}d \textbf{x} \ \partial_pV\partial^pV \ x^ax^b\,=\,-\sigma (1+\sigma)^2 \ Q_{LL}[GR],
\end{equation}
where we remind that $|\sigma|<1$ is a small dimensionless parameter \cite{Alain0A,Alain0B,Alain1,Alain2}. The second piece is more demanding because it involves second order derivative terms, 
\begin{equation}
\begin{split}
-\sigma\int_{\mathcal{M}} d\textbf{x} \ \Big[\kappa\Delta (\partial_pV\partial^pV)\Big] \ x^ax^b\,=&\,-2\epsilon\int_{\mathcal{M}} d\textbf{x} \ \Big[(\partial_p\Delta V)(\partial^pV)+(\partial_m\partial_pV)(\partial^m\partial^pV)\Big] \ x^ax^b.
\end{split}
\end{equation}
In the remaining part of this paragraph we will analyse these two contributions in a more detailed way before we will eventually assemble, in the next paragraph, the matter and field contributions. The first term is by far the easiest one and can be solved by simple partial integration,
\begin{equation}
\begin{split}
\frac{7\epsilon}{4\pi c^2G}\int_{\mathcal{M}} d\textbf{x} \ (\partial_p\Delta V)(\partial^pV) \ x^ax^b\,=&\,-7\epsilon\frac{G}{c^2}\sum_A\sum_{A\neq B} \tilde{m}_A\tilde{m}_B \ \Big[\frac{r^a_{AB} x^b_A}{r^3_{AB}}+\frac{r^b_{AB} x^a_A}{r^3_{AB}}\Big].
\end{split}
\end{equation}
This result was worked out by making use of the extended regularization prescription which was mentioned previously. In the case of a binary system this contribution finally leads to,
\begin{equation}
-\frac{7\epsilon G}{  c^2} \sum_{A=1}^2 \sum_{B\neq A=1}^2 \tilde{m}_A \tilde{m}_B \ \Big[\frac{r^a_{AB} x^b_A}{r^3_{AB}}+\frac{r^b_{AB} x^a_A}{r^3_{AB}}\Big]\,=\,-14\frac{\epsilon \tilde{\eta} m}{ c^2}  \frac{Gm}{r^3} \ r^ar^b,
\end{equation}
where we have introduced the dimensionless quantity $\tilde{\eta}=(\tilde{m}_1\tilde{m}_2)/m^2=(1+\sigma)^2\eta$. The second term is more sophisticated and therefore requires a much deeper analysis (appendix \ref{AppendixUVQuadrupole}),
\begin{equation}
\label{TwoIntMain-18.05.17}
\begin{split}
&\,\frac{7\epsilon}{4\pi c^2G}\int_{\mathcal{M}} d\textbf{x} \ (\partial_m\partial_pV)(\partial^m\partial^pV) \ x^ax^b\\
=&\,-\frac{21\epsilon}{4\pi c^2 G}\sum_A\sum_{B\neq A} \tilde{m}_A\tilde{m}_B\int_{\mathcal{M}} d\textbf{x} \ \frac{x^ax^b}{|\textbf{x}-\textbf{x}_A|^3 \ |\textbf{x}-\textbf{x}_B|^3}\\
&\,+\frac{63 \epsilon}{4\pi c^2 G}\sum_A\sum_{B\neq A} \tilde{m}_A\tilde{m}_B \int_{\mathcal{M}}d\textbf{x} \ \frac{(\textbf{x}-\textbf{x}_A)_m (\textbf{x}-\textbf{x}_A)_p}{|\textbf{x}-\textbf{x}_A|^5} \ \frac{(\textbf{x}-\textbf{x}_B)^m (\textbf{x}-\textbf{x}_B)^p}{|\textbf{x}-\textbf{x}_B|^5} \ x^ax^b
\end{split}
\end{equation}
We see that this term splits into two separate integrals \cite{WillWiseman, PatiWill1,PatiWill2,PoissonWill, WagonerWill1976,MisnerThroneWheeler}. In order to evaluate these two integrals we aim to follow the integration methods presented in \cite{Alain1,PoissonWill,PatiWill1,PatiWill2,WillWiseman},
\begin{equation}
\int_{\mathcal{M}}d\textbf{y} \ f(\textbf{y})\,=\,\int_{\mathcal{M}_y}d\textbf{y} \ f(\textbf{y})-\int_{\partial\mathcal{M}_y}\textbf{r}\cdot d\textbf{S} f(\textbf{y})+\cdots.
\end{equation}
We remind that in this context the first task is to perform the substitution $\textbf{y}:=\textbf{x}-\textbf{x}_B$ followed by a translation of the domain of integration, $\mathcal{M}\leadsto \mathcal{M}_y+\partial\mathcal{M}_y$ where the near zone domain $\mathcal{M}$ is defined by $|\textbf{x}|<\mathcal{R}$, $\mathcal{M}_y$ is defined by $|\textbf{y}|<\mathcal{R}$ and $\partial\mathcal{M}_y$ is its boundary at $y=\mathcal{R}$. It is clear that the surface integral is smaller than the volume integral by a factor $r/\mathcal{R}$ and the neglected terms are even smaller \cite{Alain0A,Alain0B,Alain1,Alain2}. In that perspective, we can split the first integral into a volume contribution $I_I^{ab}$ and a surface contribution $S_I^{ab}$, 
\begin{equation}
\begin{split}
\,&\,-\frac{21\epsilon G}{4\pi c^2 }\sum_A\sum_{B\neq A} \tilde{m}_A\tilde{m}_B\int_{\mathcal{M}} d\textbf{x} \ \frac{x^ax^b}{|\textbf{x}-\textbf{x}_A|^3 \ |\textbf{x}-\textbf{x}_B|^3}\,=\,I_I^{ab}-S_I^{ab}+\cdots,
\end{split}
\end{equation}
and the precise expressions for the two integrals are,
\begin{equation}
\begin{split}
I^{ab}_I\,=&\,-\frac{21\epsilon G}{4\pi c^2 } \sum_A\sum_{B\neq A} \tilde{m}_A\tilde{m}_B\int_{\mathcal{M}_y} d\textbf{y} \ \frac{(y+x_B)^a \ (y+x_B)^b}{|\textbf{y}-\textbf{r}_{AB}|^3 y^3},\\
S^{ab}_I\,=&\,-\frac{21\epsilon G}{4\pi c^2 }\sum_A\sum_{B\neq A} \tilde{m}_A\tilde{m}_B\int_{\partial\mathcal{M}_y} \textbf{r}\cdot d\textbf{S} \ \frac{(y+x_B)^a \ (y+x_B)^b}{|\textbf{y}-\textbf{r}_{AB}|^3 y^3}.\\
\end{split}
\end{equation}
In the framework of a binary system the four integrals of $I_I^{ab}$ translate into (appendix \ref{AppendixUVQuadrupole}):

\begin{minipage}{0.5\textwidth}
\begin{equation*}
\begin{split}
I^{ab}_{I1}\,=&\,-\frac{\epsilon\tilde{\eta}m}{c^2} \Big[\frac{49}{10}\frac{Gm}{r^3} r^ar^b+\frac{531}{30}\frac{Gm}{r} \delta^{ab}\Big],\\
I^{ab}_{I2}\,=&\,+\frac{\epsilon\tilde{\eta}m}{c^2} \frac{35}{6}\frac{Gm}{r^3} r^ar^b,\\
\end{split}
\end{equation*}
\end{minipage}
\begin{minipage}{0.5\textwidth}
\begin{equation*}
\begin{split}
I^{ab}_{I3}\,=&\,+\frac{\epsilon\tilde{\eta}m}{c^2} \frac{35}{6}\frac{Gm}{r^3} r^ar^b,\\
I^{ab}_{I4}\,=&\,-\frac{\epsilon\tilde{\eta}m}{c^2} 28 (1-2\eta) \frac{Gm}{r^3}  r^ar^b,\\
\end{split}
\end{equation*}
\end{minipage}
$\newline$

\noindent
where we have used the following dimensionless $\tilde{\eta}=(\tilde{m}_1\tilde{m}_2)/m^2=(1+\sigma)^2\eta$ quantity. The four surface integrals $S_I^{ab}$ are either proportional to the near zone cut-off parameter $\mathcal{R}$ or merely vanish. We remind that according to \cite{WillWiseman,PatiWill1,PatiWill2,PoissonWill} the results proportional to $\mathcal{R}$ will eventually be cancelled by terms coming from the wave zone and in this sense they can be freely discarded. From a technical point of view the second integral can be solved in a very similar way. There are however a couple of computational differences which we will point out and thoroughly discuss as we go through the technical details. In analogy to what has been outlined for the previous computation we need to perform a substitution of the integration variable $\textbf{y}=\textbf{x}-\textbf{x}_A$ together with a shift of the domain of integration,
\begin{equation}
\begin{split}
\sum_A\sum_{B\neq A} \tilde{m}_A\tilde{m}_B \int_{\mathcal{M}}d\textbf{x} \ \frac{(\textbf{x}-\textbf{x}_A)_m (\textbf{x}-\textbf{x}_A)_p}{|\textbf{x}-\textbf{x}_A|^5} \ \frac{(\textbf{x}-\textbf{x}_B)^m (\textbf{x}-\textbf{x}_B)^p}{|\textbf{x}-\textbf{x}_B|^5} \ x^ax^b\,=\, \frac{I_{II}^{ab}-S_{II}^{ab}}{\frac{63 \epsilon G}{4\pi c^2 }} +\cdots,
\end{split}
\end{equation}
where the precise expressions for the two contributions are the following ones,
\begin{equation}
\begin{split}
I^{ab}_{II}\,=&\,\frac{63 \epsilon G}{4\pi c^2 }\sum_A\sum_{B\neq A} \tilde{m}_A\tilde{m}_B \int_{\mathcal{M}_y} d\textbf{y}  \ \frac{(\textbf{y}-\textbf{r}_{AB})_m \ y^m \ (\textbf{y}-\textbf{r}_{AB})_p \ y^p}{|\textbf{y}-\textbf{r}_{AB}|^5 \ y^5} \ (y+x_B)^a \ (y+x_B)^b,\\
S^{ab}_{II}\,=&\,\frac{63 \epsilon G}{4\pi c^2 }\sum_A\sum_{B\neq A} \tilde{m}_A\tilde{m}_B \int_{\partial\mathcal{M}_y} \textbf{r} \cdot d\textbf{S}  \ \frac{(\textbf{y}-\textbf{r}_{AB})_m \ y^m \ (\textbf{y}-\textbf{r}_{AB})_p \ y^p}{|\textbf{y}-\textbf{r}_{AB}|^5 \ y^5} \ (y+x_B)^a \ (y+x_B)^b.
\end{split}
\end{equation}
The detailed evaluation of these two integrals can be found in the appendix \ref{AppendixUVQuadrupole} related to this subsection \cite{Alain1,Alain2}. By making use of the STF products, we obtain in the context of a binary system, the following results for the sixteen integrals of $I_{II}^{ab}$:

\begin{minipage}{0.5\textwidth}
\begin{equation*}
\begin{split}
I_{II1}^{ab}\,=&\, +\frac{\epsilon\tilde{\eta} m}{c^{2}} \ \bigg[\frac{483}{10} \frac{Gm}{ r}\delta^{ab}+\frac{63}{5} \frac{Gm}{ r^3} r^a r^b\bigg],\\
I_{II2}^{ab}\,=&\,-\frac{\epsilon \tilde{\eta} m}{ c^2} \ \frac{49}{4}\frac{G m}{ r^3} r^a r^b,\\
I_{II3}^{ab}\,=&\,-\frac{\epsilon \tilde{\eta} m}{ c^2} \ \frac{49}{4}\frac{G m}{ r^3} r^a r^b,   \\
I_{II4}^{ab}\,=&\,+\frac{\epsilon \tilde{\eta} m}{c^2} \ \frac{105}{2}\frac{G m}{r^3} (1-2\eta) r^ar^b,  \\
\end{split}
\end{equation*}
\end{minipage}
\begin{minipage}{0.5\textwidth}
\begin{equation*}
\begin{split}
I_{II5}^{ab}\,=&\,+\frac{\epsilon \tilde{\eta} m}{ c^2} \ \bigg[\frac{344}{25} \frac{Gm}{r^3} r^ar^b +\frac{179}{100} \frac{Gm}{r} \delta^{ab}\bigg],\\
I_{II6}^{ab}\,=&\, -\frac{\epsilon \tilde{\eta} m}{ c^2} \frac{1407}{25} \ \frac{Gm}{r^3} \ r^ar^b,\\
I_{II7}^{ab}\,=&\, -\frac{\epsilon \tilde{\eta} m}{ c^2} \frac{1407}{25} \ \frac{Gm}{r^3} \ r^ar^b,\\
I_{II8}^{ab}\,=&\, +\frac{\epsilon \tilde{\eta} m}{ c^2} \frac{147}{10} \ \frac{Gm}{r^3} (1-2\eta) r^ar^b,\\
\end{split}
\end{equation*}
\end{minipage}

\begin{minipage}{0.5\textwidth}
\begin{equation*}
\begin{split}
I_{II9}^{ab}\,=&\,+\frac{\epsilon\tilde{\eta} m}{c^2} \bigg[\frac{179}{50} \frac{Gm}{r} \delta^{ab}+\frac{344}{25}\frac{Gm}{r^3} r^ar^b\bigg],\\
I_{II10}^{ab}\,=&\,-\frac{\epsilon \tilde{\eta} m}{c^2}\frac{532}{25} \frac{Gm}{r^3} r^ar^b,\\
I_{II11}^{ab}\,=&\,-\frac{\epsilon \tilde{\eta} m}{c^2}\frac{532}{25} \frac{Gm}{r} r^ar^b,\\
I_{II12}^{ab}\,=&\,+\frac{\epsilon\tilde{\eta} m}{c^2}\frac{147}{10} \frac{Gm}{r^3}r^ar^b (1-2\eta),
\end{split}
\end{equation*}
\end{minipage}
\begin{minipage}{0.5\textwidth}
\begin{equation*}
\begin{split}
I_{II13}^{ab}\,=&\,+\frac{\epsilon \tilde{\eta} m}{c^2} \bigg[\frac{7489}{350} \frac{Gm}{r^3} r^ar^b+\frac{7397}{1050} \frac{Gm}{r} \delta^{ab}\bigg],\\
I_{II14}^{ab}\,=&\,-\frac{\epsilon \tilde{\eta} m}{c^2} \frac{1791}{175} \frac{Gm}{r^3} r^ar^b,\\
I_{II15}^{ab}\,=&\,-\frac{\epsilon \tilde{\eta} m}{c^2} \frac{1791}{175} \frac{Gm}{r^3} r^ar^b,\\
I_{II16}^{ab}\,=&\,+\frac{\epsilon \tilde{\eta} m}{c^2}\frac{32}{5} \frac{Gm}{r^3} (1-2\eta) r^ar^b.
\end{split}
\end{equation*}
\end{minipage}
$\newline$

\noindent
The third contribution of the exponential operator, $-\frac{\epsilon\kappa}{2} \int_{\mathcal{M}}d\textbf{x} \ \Delta^2 \ \partial_pV\partial^pV$, is more strongly suppressed than the previous ones. We saw, by construction of the model, that the parameters are, in the International System of Units (SI), smaller than one, $|\sigma| <1, \ \kappa <1 \ (\epsilon=\sigma\kappa)$. We will therefore not review these highly suppressed contributions in this manuscript \cite{Alain1,Alain2}. With this we can finally sum up all the different near zone field contributions, for the case of a binary system, into one single expression,
\begin{equation}
Q^{ab}_{LL}\,=\,(1+\sigma)^2(1-\sigma) Q^{ab}_{LL}[GR]+Q^{ab}_{LL}[NL]+\mathcal{O}(c^{-4}),
\end{equation}
where the 1.5 post-Newtonian general relativistic and nonlocal field contributions are of the following forms,
\begin{gather}
Q^{ab}_{LL}[GR]\,=\,-\frac{7}{2}(1-2\eta) \frac{Gm}{c^2r},\\
Q^{ab}_{LL}[NL]\,=\,(1+\sigma)^2\frac{\epsilon\eta m}{c^2} \Big[A\frac{Gm}{r^3}r^ar^b+B\frac{Gm}{r^3}(1-2\eta) r^ar^b +C\frac{Gm}{r}\delta^{ab}\Big].
\end{gather}
Here we have for the three numerical coefficients, $A=-\frac{10231}{70}$, $B=+\frac{603}{10}$ and $C=+\frac{7729}{210}$. Additional computational details can be withdrawn from the appendix \ref{AppendixUVQuadrupole} related to this subsection.

\subsubsection{The nonlocal UV radiative quadrupole moment:}
In this paragraph we aim to assemble the matter and field contributions in order to obtain the total 1.5 post-Newtonian near zone radiative quadrupole moment in the context of the UV modified theory of gravity,
\begin{equation*}
Q^{ab}\,=\,Q^{ab}[GR]+Q^{ab}[NL]+\mathcal{O}(c^{-4}).
\end{equation*}
Taking into account the way in which the intermediate results for the matter and field contributions have been outlined in the previous two paragraphs of this subsection, it is rather obvious that we can split the final result in a general relativistic component followed by a nonlocal correction term,
\begin{equation*}
\label{QM-27.11.17}
\begin{split}
Q^{ab}[GR]\,&=\,+\eta m \Big[1+\frac{1-3\eta}{2}\frac{v^2}{c^2}-\frac{1-2\eta}{2}\frac{Gm}{c^2r}\Big]r^ar^b,\\
Q^{ab}[NL]\,&=\,-\frac{\eta m}{c^2}\bigg[z(\sigma) \ (1-2\eta)\frac{Gm}{r} r^ar^b+\Big( g(\epsilon,\kappa)- C \ \frac{k^2(\epsilon,\kappa)}{\epsilon} \ \Big) \frac{Gm}{r}\delta^{ab}\bigg]\\
& \quad \  -\frac{\eta m}{c^2}\bigg[\Big(g(\epsilon,\kappa)- \big[A- B \ (1-2\eta)\big] \ \frac{k^2(\epsilon,\kappa)}{\epsilon} \  \Big) \frac{Gm}{r^3}r^ar^b\bigg].
\end{split}
\end{equation*}
For clarity and comprehensibility reasons we would like to remind that we have,

\begin{minipage}{0.33\textwidth}
\begin{equation*}
\begin{split}
A\,&=\,-\frac{10231}{70},\\
\epsilon&=\sigma \kappa,\\
m&=m_1+m_2,
\end{split}
\end{equation*}
\end{minipage}
\begin{minipage}{0.33\textwidth}
\begin{equation*}
\begin{split}
B\,&=\,+\frac{603}{10},\\
g(\epsilon,\kappa)\,&=\,8\epsilon \ \frac{1+\sigma}{1-\sigma},\\
z(\sigma)&=\sigma/2+\sigma^2+\sigma^3,
\end{split}
\end{equation*}
\end{minipage}
\begin{minipage}{0.33\textwidth}
\begin{equation*}
\begin{split}
C\,&=\,+\frac{7729}{210},\\
k(\epsilon,\kappa)\,&=\,\epsilon \ (1+\sigma)^2,\\
\eta&= \frac{m_1 m_2}{m^2}.\\
\end{split}
\end{equation*}
\end{minipage}
$\newline$
It should be noticed that in the limit of vanishing modification parameters ($\sigma$, $\kappa$) we recover the well known general relativistic term \cite{WillWiseman,PatiWill1,PoissonWill} at the 1.5 post-Newtonian order of accuracy,
\begin{equation*}
\lim_{\sigma\rightarrow 0,\kappa\rightarrow 0} Q^{ab}=Q^{ab}[GR]+\mathcal{O}(c^{-4}).
\end{equation*}
In summary we saw throughout the different subsections of this chapter that we are able to recover, for any of the quantities that we computed in the context of the nonlocal field theory of gravity, the general relativistic result in the particular limits mentioned above. 

\section{Conclusion}
\label{Conclusion}

We observe that corrections originating from nonlocality set in at the 1.0 post-Newtonian order of accuracy. We saw in table \ref{TableStrengths} that for astrophysical systems the deviation in the coupling strength compared to standard theory of gravity is of the order, $|G-G_\kappa(\lambda_D)|\sim 10^{-4}$. We remind from \cite{Alain0B} that findings obtained in the framework of calculations performed previously in the context of possible deviations of the perihelion precession of Mercury led to a value for the dimensionless UV parameter, $|\sigma|\leq 9.3\cdot 10^{-4}$. It is interesting to notice that the leading order deviation term in equation \eqref{QM-27.11.17} is proportional to $z(\sigma)\propto\sigma/2\sim 10^{-4}$, which coincides pretty well with the result obtained in table \ref{TableStrengths}. Moreover it should be noticed that the term proportional to $\delta^{ab}$ disappears after the use of the TT-projection \cite{WillWiseman,PoissonWill,PatiWill1,PatiWill2}. For rather large $\kappa$ values and short orbital separation distances ($r\rightarrow 0$) the contribution in $^{ab}[NL]$ proportional to $g(\epsilon,\kappa)\ r^{-3}$ could become the dominant contribution. In an binary system merging scenario the term proportional to $r^{-3}$ will eventually become the dominant piece as it will diverge more rapidly than the remaining contributions.

\begin{appendix}
\section{The effective UV radiative quadrupole moment}
\label{AppendixUVQuadrupole}
Technical details about the derivation of the different components of the UV radiative quadrupole moment are presented in the remaining part of this appendix-subsection. 
\subsubsection{The matter contribution $Q_m^{ab}$}
We will provide in this appendix-subsection further computational details regarding the different 1.5 post-Newtonian near zone matter contributions to the radiative quadrupole moment outlined in subsection \ref{UVQuadrupole}. The first contribution gives rise to the usual general relativistic matter-term $Q^{ab}_m[GR]$ plus an additional term that disappears in the limit of vanishing $\sigma$,
\begin{equation}
\begin{split}
Q^{ab}_{\mathcal{B}_{1}+\mathcal{B}h}\,=\,c^{-2}\int_{\mathcal{M}}d\textbf{x} \ \Big[\mathcal{B}^{00}_{1}+\mathcal{B}^{00}h^{00}\Big]\ x^ax^b\,
=&\,\int_{\mathcal{M}}d\textbf{x} \ \sum_A \Big[1+\frac{v^2_A}{2c^2}+3\frac{G}{c^2} \sum_{B\neq A} \frac{\tilde{m}_B}{|\textbf{x}-\textbf{x}_B|}\Big] \ \delta(\textbf{x}-\textbf{x}_A)+\mathcal{O}(c^{-4})\\
=&\,Q^{ab}_m[GR]+3\sigma \frac{G}{c^2} \sum_A\sum_{B\neq A} \frac{m_Am_B}{|\textbf{x}_A-\textbf{x}_B|} \ x^a_A x^b_A+\mathcal{O}(c^{-4}),
\end{split}
\end{equation}
where $Q^{ab}_m[GR]=\sum_Am_A\Big(1+\frac{v^2_A}{2c^2}+3\frac{G}{c^2}\sum_{B\neq A} \frac{m_B}{|\textbf{x}_A-\textbf{x}_B|}\Big) \ x^a_A x^b_A$.
It should be noticed from what was said in the main body of this manuscript that the curvature energy-momentum contribution \cite{Alain0A,Alain0B,Alain1} $\mathcal{B}_2^{\alpha\beta}$ can be expanded in the following way,
\begin{equation*}
\begin{split}
\mathcal{B}^{\alpha\beta}_2\,=&\,\epsilon \ \Big[\sum_{n=0}^\infty \sigma^n e^{(n+1)\kappa \Delta} \Big] \ \Big[-\frac{h^{00}}{2}\Delta\Big] \  \Big[ \sum_A m_A v^\alpha_A v^\beta_A \delta(\textbf{y}-\textbf{x}_A)\Big]+\mathcal{O}(c^{-4})\\
=&\,-\frac{\epsilon}{2} \frac{1}{1-\sigma} \sum_A m_A v_A^\alpha v^\beta_A \  \Bigg[h^{00} \Big(\Delta \delta(\textbf{y}-\textbf{x}_A)\Big)\Bigg]   \\
 &\,-\frac{\epsilon}{2} \frac{\kappa}{(1-\sigma)^2} \sum_A m_A v_A^\alpha v^\beta_A \Delta \Bigg[h^{00} \Big(\Delta \delta(\textbf{y}-\textbf{x}_A)\Big)\Bigg] \\
 &\,-\frac{\epsilon}{2} \sum_A m_A v_A^\alpha v^\beta_A \   \sum_{n=0}^{+\infty} \sigma^n \ \sum_{m=2}^{+\infty} \frac{[(n+1)\kappa]^m}{m!} \Delta^m \Bigg[h^{00} \Big(\Delta \delta(\textbf{y}-\textbf{x}_A)\Big)\Bigg]+\mathcal{O}(c^{-4})
\end{split}
\end{equation*}
We can use this term inside the quadrupole integral and we obtain obtain,
\begin{equation}
\begin{split}
Q^{ab}_{\mathcal{B}_{A5}}\,=\,c^{-2} \int_{\mathcal{M}}d\textbf{x} \ \mathcal{B}^{00}_5 \ x^ax^b\,
=&\,\epsilon \int_{\mathcal{M}}d\textbf{x}  \ \Big[\sum_{n=0}^\infty \sigma^n e^{(n+1)\kappa \Delta} \Big] \ \Big[-\frac{h^{00}}{2}\Delta\Big] \  \Big[ \sum_A m_A  \delta(\textbf{y}-\textbf{x}_A)\Big] x^ax^b+\mathcal{O}(c^{-4})\\
=&\,\epsilon \frac{1+\sigma}{1-\sigma} \frac{4G}{c^2} \sum_A\sum_{B\neq A} m_Am_B \ \bigg[\frac{(\textbf{x}_A-\textbf{x}_B)^b}{|\textbf{x}_A-\textbf{x}_B|^3} x_A^{a} +\frac{(\textbf{x}_A-\textbf{x}_B)^a}{|\textbf{x}_A-\textbf{x}_B|^3} x_A^b-\frac{\delta^{ab}}{|\textbf{x}_A-\textbf{x}_B|}\bigg]+\mathcal{O}(c^{-4}).\\
\end{split}
\end{equation}
It should be noticed that although the penultimate line of computation contains a sum of infinitely many derivatives, the result at 1.5 post-Newtonian order of accuracy is finite. This is due to the fact that only the lowest derivative terms in this line of computation eventually contribute to the final result. Higher order derivative terms will lead, after partial integration(s), to contributions that are proportional to either one of the following two relations or to both relations at the same time,
\begin{equation}
\label{cond-17.05.17}
\begin{split}
\sum_A\sum_{B\neq A}m_A \tilde{m}_B \ \nabla^q\delta(\textbf{x}_A-\textbf{x}_B)\,=\,0,\quad \forall q\geq 0,\quad\nabla^q [x^ax^b]\,=\,0, \quad \forall q\geq 3.
\end{split}
\end{equation}
Surface terms arising from partial integration can be freely discarded in the near zone domain $\mathcal{M}: \ \textbf{x}<\mathcal{R}$,
\begin{equation}
\label{surface-17.05.17}
\int_{\partial \mathcal{M}} dS^p \ \partial_p h^{00}\frac{\delta(\mathbf{y}-\mathbf{r}_A)}{|\mathbf{x}-\mathbf{y}|}  \propto \delta(\mathcal{R}-|\mathbf{x}_A|).
\end{equation}
The derivative contribution $D^{\alpha\beta}$ at the 1.5 post-Newtonian order of accuracy becomes,
\begin{equation}
\begin{split}
Q^{ab}_{D}\,=\,-\frac{\sigma}{c^2} \int_{\mathcal{M}}d\textbf{x} \ D^{00} \ x^ax^b\,=&\,-\sigma\sum_A m_A \ \mathcal{S}(\sigma,\kappa) \int_{\mathcal{M}}d\textbf{x} \ \Big[ \nabla^{2p+2n-m} \delta(\textbf{x}-\textbf{x}_A)\Big]\Big[ \nabla^m h^{00}\Big] \ x^ax^b +\mathcal{O}(c^{-4})\\
=&\,+\epsilon\frac{1+\sigma}{1-\sigma} \frac{8G}{c^2} \ \sum_A \sum_{B\neq A}m_A  m_B \ \bigg[ \frac{(\textbf{x}_B-\textbf{x}_A)^b}{|\textbf{x}_A-\textbf{x}_B|^3} x_A^a+\ \frac{(\textbf{x}_B-\textbf{x}_A)^a}{|\textbf{x}_A-\textbf{x}_B|^3} x_A^b\bigg]+\mathcal{O}(c^{-4}),
\end{split}
\end{equation}
where $S(\sigma,\kappa)$ was outlined in \cite{Alain0A,Alain1}. It should be noticed that, although $D^{\alpha\beta}$ involves infinitely many terms, this result is the precise result at the 1.5 post-Newtonian order of accuracy. This is due to the fact that only the lowest derivative terms  in the penultimate line of computation above contribute to the final result. The lowest derivative term acting on the Dirac distribution is obtained for $n=1, \ m=1, \ p=0$. Higher order derivative terms will lead, after partial integration(s), to contributions that are proportional to either one or to both relations at the same time outlined in equation \eqref{cond-17.05.17}. Surface terms arising from partial integration can be freely discarded in the near zone domain as we already argued in equation \eqref{surface-17.05.17}. This allows us to write down the complete near zone  matter radiative  quadrupole moment at the 1.5 post-Newtonian order of accuracy,
\begin{equation}
\begin{split}
Q^{ab}_m\,=&\,Q^{ab}_{m}[GR]-4\epsilon\frac{1+\sigma}{1-\sigma} \ \frac{ G}{ c^2}\ \sum_A \sum_{B\neq A}m_A m_B \ \bigg[\frac{r_{AB}^ar^b_A}{r_{AB}^3}+\frac{r_{AB}^b r^a_A}{r_{AB}^3}+\frac{\delta^{ab}}{r_{AB}}\bigg]+3\sigma \sum_A\sum_{B\neq A}\frac{r^a_Ar^b_A}{r_{AB}}+\mathcal{O}(c^{-4}),
\end{split}
\end{equation}
for an $n$-body system. After having worked out the matter contribution we will turn now to the field contribution.
\subsubsection{The field contribution $Q_{LL}^{ab}$}
In the quest of working out the 1.5 post-Newtonian field contribution to the near zone radiative quadrupole moment $Q^{ab}_{LL}$ we have,
\begin{equation}
\begin{split}
Q^{ab}_{LL}\,=&\,c^{-2} \int_{\mathcal{M}} d\textbf{x} \  N^{00}_{LL} \ x^ax^b\,=\,-\frac{7}{8\pi c^2G} \int_{\mathcal{M}} d\textbf{x} \ \Big[1-\sigma e^{\kappa\Delta}\Big] \ \Big[\partial_pV\partial^pV\Big] \ x^ax^b+\mathcal{O}(c^{-4})\\
\end{split}
\end{equation}
We adopt the strategy to analyse the two remaining contributions independently one after the other. The first term leads to a contribution that is proportional to the usual 1.5 post-Newtonian general relativistic \cite{WillWiseman,PatiWill1,PoissonWill} contribution,
\begin{equation}
\begin{split}
-\frac{7}{8\pi c^2G} \int_{\mathcal{M}} d\textbf{x} \ \partial_pV\partial^pV \ x^ax^b\,=&\,-\frac{7}{8\pi c^2G} \bigg[\int_{\mathcal{M}} d\textbf{x} \ \partial_p(V\partial^pV \ x^ax^b)-\int_{\mathcal{M}} d\textbf{x} \ V(\nabla^2 V) \ x^ax^b\bigg]\\
&\,-\frac{7}{8\pi c^2 G}\bigg[\int_{\mathcal{M}} d\textbf{x} \ V^2\delta^{ab}-\frac{1}{2}\int_{\mathcal{M}} d\textbf{x} \ \big[\partial^a(V^2x^b+\partial^b(V^2x^a)\big]\bigg]\\
=&\,-\frac{7G}{2c^2}\sum_A\sum_{B\neq A} \frac{\tilde{m}_A\tilde{m}_B}{r_{AB}}r^a_Ar^b_A-\frac{7}{2Gc^2}\Big[\mathcal{R}^4 \langle V\partial_pV N^aN^bN^p\rangle-\mathcal{R}^3\langle V^2N^aN^b\rangle\Big]
\end{split}
\end{equation}
It should be noticed that the result displayed above is correct only with respect to the fact that the term $V^2\delta^{ab}$ has been discarded as it will not survive a TT-projection \cite{WillWiseman,PoissonWill,PatiWill1}. Moreover it can be shown that the last two quantities are both $\mathcal{R}$ dependent. By taking into account these various considerations, it can easily be seen that the integral is proportional to the usual 1.5 post-Newtonian result $(1+\sigma)^2 Q^{ab}_{LL}$. For completeness we wish to present the binary system result before we move on to analyse the next contribution,
\begin{equation}
\begin{split}
Q^{ab}_{LL}[GR]\,=\,-\frac{7G}{2c^2}\sum_{A=1}^2\sum_{B\neq A=1}^2 \frac{\tilde{m}_A\tilde{m}_B}{r_{AB}}r^a_Ar^b_A\,=\, -\frac{7}{2c^2} \eta m \ (1-2\eta) \frac{Gm}{r} \ r^ar^b.
\end{split}
\end{equation}
The second term entails, through the exponential differential operator, a sum of infinitely many derivatives,
\begin{equation}
\begin{split}
-\sigma\int_{\mathcal{M}} d\textbf{x} \ \Big[ e^{\kappa\Delta} \ \partial_pV\partial^pV\Big] \ x^a x^b\,=&\,-\sigma\int_{\mathcal{M}} d\textbf{x} \ \Big[ (1+\kappa \Delta+\frac{\kappa^2}{2}\Delta^2) \ \partial_pV\partial^pV\Big] \ x^a x^b+\cdots.
\end{split}
\end{equation}
However a careful analysis shows that higher order derivative terms will lead, after partial integration(s), to contributions that are proportional to either one of or to both relations at the same time displayed in equation \eqref{cond-17.05.17}. Surface terms arising from partial integration can be freely discarded in the near zone domain because of equation \eqref{surface-17.05.17}. The remaining threes contributions of this second term have to be analysed separately. A closer look reveals that the first one is very similar to the one which has already been worked out in this subsection and leads to another contribution which is directly proportional to the 1.5 post-Newtonian general relativistic field contribution,$-\sigma \int_{\mathcal{M}}d \textbf{x} \ \partial_pV\partial^pV \ x^ax^b=-\sigma (1+\sigma)^2 \ Q_{LL}[GR]$. The second term is more demanding because, in addition to the complexity encountered so far, it involves second order derivatives, 
\begin{equation}
\begin{split}
-\sigma\int_{\mathcal{M}} d\textbf{x} \ \Big[\kappa\Delta (\partial_pV\partial^pV)\Big] \ x^ax^b\,=&\,-2\epsilon\int_{\mathcal{M}} d\textbf{x} \ \Big[(\partial_p\Delta V)(\partial^pV)+(\partial_m\partial_pV)(\partial^m\partial^pV)\Big] \ x^ax^b.
\end{split}
\end{equation}
In the remaining part of this subsection we will analyze these two contributions in a more detailed way. The first one is by far the less demanding one and can be solved by simple partial integration,
\begin{equation}
\begin{split}
\frac{7\epsilon}{4\pi c^2G}\int_{\mathcal{M}} d\textbf{x} \ (\partial_p\Delta V)(\partial^pV) \ x^ax^b\,
=&\,\frac{7\epsilon}{Gc^2}\sum_A \tilde{m}_A \int_{\mathcal{M}} d\textbf{x} \  \delta(\textbf{x}-\textbf{x}_A) \ \Big[\Delta V x^ax^b+(\partial^aV) x^b+(\partial^bV)x^a\Big]\\
=&\,-7\epsilon\frac{G}{c^2}\sum_A\sum_{A\neq B} \tilde{m}_A\tilde{m}_B \ \Big[\frac{r^a_{AB} x^b_A}{r^3_{AB}}+\frac{r^b_{AB} x^a_A}{r^3_{AB}}\Big].
\end{split}
\end{equation}
It should be noticed that, for later convenience, we aimed to take care of the precise coefficients in front of the integral. By making use of the regularization prescription we were able to discard the first term in this integral, $\sum_A\sum_{B\neq A} \delta(\textbf{x}_A-\textbf{x}_B)=0$. For the case of a binary system this contribution leads to the following result,
\begin{equation}
 -\frac{7\epsilon G}{  c^2} \sum_{A=1}^2 \sum_{B\neq A=1}^2 \tilde{m}_A \tilde{m}_B \ \Big[\frac{r^a_{AB} x^b_A}{r^3_{AB}}+\frac{r^b_{AB} x^a_A}{r^3_{AB}}\Big]\,=\,-14\frac{\epsilon \tilde{\eta} m}{ c^2}  \frac{Gm}{r^3} \ r^ar^b,
\end{equation}
where we have introduced the following dimensionless $\tilde{\eta}=(\tilde{m}_1\tilde{m}_2)/m^2=(1+\sigma)^2\eta$. The second term is more sophisticated and therefore requires a much deeper analysis,
\begin{equation}
\label{TwoInt-18.05.17}
\begin{split}
\frac{7\epsilon}{4\pi c^2G}\int_{\mathcal{M}} d\textbf{x} \ (\partial_m\partial_pV)(\partial^m\partial^pV) \ x^ax^b\,
=&\,\frac{7\epsilon}{4\pi c^2G}\sum_A\sum_{B\neq A} \tilde{m}_A\tilde{m}_B\int_{\mathcal{M}} d\textbf{x} \ \bigg[\frac{-\delta_{mp}}{|\textbf{x}-\textbf{x}_A|^3}+3\frac{(\textbf{x}-\textbf{x}_A)_p \ (\textbf{x}-\textbf{x}_A)_m}{|\textbf{x}-\textbf{x}_A|^5}\bigg]\\
&\hspace{2cm} \bigg[\frac{-\delta^{mp}}{|\textbf{x}-\textbf{x}_B|^3}+3\frac{(\textbf{x}-\textbf{x}_B)^p \ (\textbf{x}-\textbf{x}_B)^m}{|\textbf{x}-\textbf{x}_B|^5}\bigg] \ x^ax^b\\
=&\,-\frac{21\epsilon}{4\pi c^2 G}\sum_A\sum_{B\neq A} \tilde{m}_A\tilde{m}_B\int_{\mathcal{M}} d\textbf{x} \ \frac{x^ax^b}{|\textbf{x}-\textbf{x}_A|^3 \ |\textbf{x}-\textbf{x}_B|^3}\\
&\,+\frac{63 \epsilon}{4\pi c^2 G}\sum_A\sum_{B\neq A} \tilde{m}_A\tilde{m}_B \int_{\mathcal{M}}d\textbf{x} \ \frac{(\textbf{x}-\textbf{x}_A)_m (\textbf{x}-\textbf{x}_A)_p}{|\textbf{x}-\textbf{x}_A|^5} \ \frac{(\textbf{x}-\textbf{x}_B)^m (\textbf{x}-\textbf{x}_B)^p}{|\textbf{x}-\textbf{x}_B|^5} \ x^ax^b
\end{split}
\end{equation}
We see that this term itself splits into two separate integrals. In this regard we can split the first integral in a volume contribution $I_I^{ab}$ and a surface contribution $S_I^{ab}$, 
\begin{equation}
\begin{split}
\,&\,-\frac{21\epsilon G}{4\pi c^2 }\sum_A\sum_{B\neq A} \tilde{m}_A\tilde{m}_B\int_{\mathcal{M}} d\textbf{x} \ \frac{x^ax^b}{|\textbf{x}-\textbf{x}_A|^3 \ |\textbf{x}-\textbf{x}_B|^3}\,=\,I_I^{ab}-S_I^{ab}+\cdots,
\end{split}
\end{equation}
where the precise expressions for the two contributions are the following ones,
\begin{equation}
\begin{split}
I^{ab}_I\,=&\,-\frac{21\epsilon G}{4\pi c^2 } \sum_A\sum_{B\neq A} \tilde{m}_A\tilde{m}_B\int_{\mathcal{M}_y} d\textbf{y} \ \frac{(y+x_B)^a \ (y+x_B)^b}{|\textbf{y}-\textbf{r}_{AB}|^3 y^3},\\
S^{ab}_I\,=&\,-\frac{21\epsilon G}{4\pi c^2 }\sum_A\sum_{B\neq A} \tilde{m}_A\tilde{m}_B\int_{\partial\mathcal{M}_y} \textbf{r}\cdot d\textbf{S} \ \frac{(y+x_B)^a \ (y+x_B)^b}{|\textbf{y}-\textbf{r}_{AB}|^3 y^3}.\\
\end{split}
\end{equation}
For clarity reasons we aim to introduce a notation which will allow us to present the important intermediate results in a more concise way,
\begin{equation}
\begin{split}
I^{ab}_I\,=\,-\frac{21\epsilon G}{4\pi c^2 } \sum_A\sum_{B\neq A} \tilde{m}_A\tilde{m}_B \ J^{ab}_I, \quad \ \ S^{ab}_I\,=\,-\frac{21\epsilon G}{4\pi c^2 } \sum_A\sum_{B\neq A} \tilde{m}_A\tilde{m}_B \ W_I^{ab}.
\end{split}
\end{equation}
Before we work out the total eight contributions of $I_{I}^{ab}$ and $S_{I}^{ab}$ we need to set in place a couple of techniques which will be used during the computations. We will basically rely on the concepts presented in \cite{WillWiseman,PatiWill1,PatiWill2,PoissonWill} and adjust them to our needs if necessary. The addition theorem for spherical harmonics, displayed in equation \eqref{SH-28.03.17}, will be used extensively during the evaluation of the integral $I_I^{ab}$, 
\begin{equation}
\begin{split}
\frac{1}{|\textbf{y}-\textbf{r}_{AB}|^3}\,=\,\frac{1}{|y|^2} \Big[1-\frac{1}{|1-|\frac{\textbf{r}_{AB}}{y}|^2|}\Big] \ \frac{1}{|\textbf{y}-\textbf{r}_{AB}|}\,
=&\,\sum_{l=0}^{+\infty}\sum_{m=-l}^{l} \frac{4\pi}{2l+1}\frac{r^l_<}{r_>^{l+3}} Y^*_{lm}(\textbf{n}_{AB})Y^{lm}(\textbf{N})+\mathcal{O}(|\textbf{r}_{AB}/y|),
\end{split}
\end{equation}
in which $r_<:=\text{min}(y,r_{AB}), \ r_>=\text{max}(y,r_{AB}), \ \textbf{N}:=\textbf{y}/y$, and $\textbf{n}_{AB}:=\textbf{r}_{AB}/r_{AB}$. Another important preliminary result has to be mentioned before we come to the precise computations of the four integrals of $I_{I}^{ab}$, $\sum_{m=-l}^lY^*_{lm}(\textbf{n}_{AB})\int Y_{lm}(\textbf{N})N^{\langle L'\rangle} d\Omega=\delta_{l'l} \ n^{\langle L\rangle}_{AB}$. Furthermore we define the following quantity,
\begin{equation}
\begin{split}
N(l,n)\,=\, \int_0^{\mathcal{R}} dy \ y^n \frac{r^l_<}{r^{l+3}_>}\,=\,\frac{-2l-3}{(n+l+1)(n-l-2)}r^{n-2}.
\end{split}
\end{equation}
Last but not least we have to mention that we profusely used products of the so called symmetric tracefree tensors, or STF tensors for short. Taking into account the various preliminary results presented above, we finally obtain for the four integrals of $J^{ab}_I$:

\hspace{-1cm}\begin{minipage}{0.6\textwidth}
\begin{equation*}
\begin{split}
J^{ab}_{I1}\,=\,\int_{\mathcal{M}_y} \frac{d\textbf{y}}{|\textbf{y}-\textbf{r}_{AB}|^3} \frac{y^ay^b}{y^3}\,=&\,\frac{7\pi}{15}\frac{r^a_{AB}r^b_{AB}}{r^3_{AB}}+\frac{83\pi}{45}\frac{\delta^{ab}}{r_{AB}},\\
J^{ab}_{I2}\,=\,\int_{\mathcal{M}_y} \frac{d\textbf{y}}{|\textbf{y}-\textbf{r}_{AB}|^3} \frac{y^ar_B^b}{y^3}\,=&\,\frac{10\pi}{9}\frac{r^a_{AB}r^b_B}{r^3_{AB}},\\
\end{split}
\end{equation*}
\end{minipage}
\begin{minipage}{0.4\textwidth}
\begin{equation*}
\begin{split}
J^{ab}_{I3}\,=\,\int_{\mathcal{M}_y} \frac{d\textbf{y}}{|\textbf{y}-\textbf{r}_{AB}|^3} \frac{r_B^ay^b}{y^3}\,=&\,\frac{10\pi}{9}\frac{r^a_Br^b_{AB}}{r^3_{AB}},\\
J^{ab}_{I4}\,=\,\int_{\mathcal{M}_y} \frac{d\textbf{y}}{|\textbf{y}-\textbf{r}_{AB}|^3} \frac{r_B^ar_B^b}{y^3}\,=&\,\frac{16\pi}{3}\frac{r^a_Br^b_{B}}{r^3_{AB}}.\\
\end{split}
\end{equation*}
\end{minipage}

$\newline$

It should be noticed that, in order to avoid logarithmic divergences, the last result has been worked out by making use of $O\,:=\,\int_0^\mathcal{R} y^n \frac{r_<^l}{r_>^{l+4}}dy=\frac{-2l-4}{(n+l+1)(n-l-3)} r^{n-3}_{AB}$. After this we can look at the four surface integrals from $W_I^{ab}$:

\hspace{-0.4cm}\begin{minipage}{0.5\textwidth}
\begin{equation*}
\begin{split}
W_{I1}^{ab}\,=\,\int_{\partial\mathcal{M}_y} \frac{\textbf{r}\cdot d\textbf{S}}{|\textbf{y}-\textbf{r}_{AB}|^3} \frac{y^ay^b}{y^3}\,=&\,0, \\
W_{I2}^{ab}\,=\,\int_{\partial\mathcal{M}_y} \frac{\textbf{r}\cdot d\textbf{S}}{|\textbf{y}-\textbf{r}_{AB}|^3} \frac{r_B^ay^b}{y^3}\,\propto&\, \mathcal{R}^{-3},
\end{split}
\end{equation*}
\end{minipage}
\begin{minipage}{0.5\textwidth}
\begin{equation*}
\begin{split}
W_{I3}^{ab}\,=\,\int_{\partial\mathcal{M}_y} \frac{\textbf{r}\cdot d\textbf{S}}{|\textbf{y}-\textbf{r}_{AB}|^3} \frac{y^ay^b}{y^3}\,\propto&\, \mathcal{R}^{-3},\\
W_{I4}^{ab}\,=\,\int_{\partial\mathcal{M}_y} \frac{\textbf{r}\cdot d\textbf{S}}{|\textbf{y}-\textbf{r}_{AB}|^3} \frac{r_B^ay^b}{y^3}\,=&\,0.
\end{split}
\end{equation*}
\end{minipage}
$\newline$

We see that the integrals either vanish or that they are proportional to the near zone cut-off parameter $\mathcal{R}$. For a binary system the four integrals of $I_I^{ab}$ are outlined in the main text of this tarticle in subsection \ref{UVQuadrupole}. From a technical point of view the second integral in equation \eqref{TwoInt-18.05.17} (equation \eqref{TwoIntMain-18.05.17} in subsection \ref{UVQuadrupole}) can be solved in a very similar way. There are however a couple of computational differences which will be pointed out as we go through the computational details. In analogy to what has been outlined for the previous computation we need to perform a substitution of the integration variable $\textbf{y}=\textbf{x}-\textbf{x}_A$ together with a shift of the domain of integration,
\begin{equation*}
\begin{split}
&\,\frac{63 \epsilon G}{4\pi c^2 }\sum_A\sum_{B\neq A} \tilde{m}_A\tilde{m}_B \int_{\mathcal{M}}d\textbf{x} \ \frac{(\textbf{x}-\textbf{x}_A)_m (\textbf{x}-\textbf{x}_A)_p}{|\textbf{x}-\textbf{x}_A|^5} \ \frac{(\textbf{x}-\textbf{x}_B)^m (\textbf{x}-\textbf{x}_B)^p}{|\textbf{x}-\textbf{x}_B|^5} \ x^ax^b\,=\,I_{II}^{ab}-S_{II}^{ab}+\cdots
\end{split}
\end{equation*}
where the precise expressions for the two contributions are the following ones,
\begin{equation*}
\begin{split}
I^{ab}_{II}\,=&\,\frac{63 \epsilon G}{4\pi c^2 }\sum_A\sum_{B\neq A} \tilde{m}_A\tilde{m}_B \int_{\mathcal{M}_y} d\textbf{y}  \ \frac{(\textbf{y}-\textbf{r}_{AB})_m \ y^m \ (\textbf{y}-\textbf{r}_{AB})_p \ y^p}{|\textbf{y}-\textbf{r}_{AB}|^5 \ y^5} \ (y+x_B)^a \ (y+x_B)^b,\\
S^{ab}_{II}\,=&\,\frac{63 \epsilon G}{4\pi c^2 }\sum_A\sum_{B\neq A} \tilde{m}_A\tilde{m}_B \int_{\partial\mathcal{M}_y} \textbf{r} \cdot d\textbf{S}  \ \frac{(\textbf{y}-\textbf{r}_{AB})_m \ y^m \ (\textbf{y}-\textbf{r}_{AB})_p \ y^p}{|\textbf{y}-\textbf{r}_{AB}|^5 \ y^5} \ (y+x_B)^a \ (y+x_B)^b.
\end{split}
\end{equation*}
Before we aim to work out the total thirty-two contributions of $I_{II}^{ab}$ and $S_{II}^{ab}$ we need to set in place a couple of techniques which will be used during the computations. In analogy to what has been presented before, the addition theorem for spherical harmonics this time reads as follows,
\begin{equation}
\begin{split}
\frac{1}{|\textbf{y}-\textbf{r}_{AB}|^5}\,=&\,\frac{1}{|y|^4} \Big[1-\frac{1}{|1-|\frac{\textbf{r}_{AB}}{y}|^4|}\Big] \ \frac{1}{|\textbf{y}-\textbf{r}_{AB}|}\\
=&\,\sum_{l=0}^{+\infty}\sum_{m=-l}^{l} \frac{4\pi}{2l+1}\frac{r^l_<}{r_>^{l+5}} Y^*_{lm}(\textbf{n}_{AB})Y^{lm}(\textbf{N})+\mathcal{O}(|\textbf{r}_{AB}/y|),
\end{split}
\end{equation}
in which again $r_<:=\text{min}(y,r_{AB}), \ r_>=\text{max}(y,r_{AB}), \ \textbf{N}:=\textbf{y}/y$, and $\textbf{n}_{AB}:=\textbf{r}_{AB}/r_{AB}$. Here again \cite{WillWiseman, PatiWill1,PatiWill2,PoissonWill}, in analogy to what was said previously, we define the following relation,
\begin{equation}
\begin{split}
L(l,n)\,=\, \int_0^{\mathcal{R}} dy \ y^n \frac{r^l_<}{r^{l+5}_>}\,=\,\frac{-2l-5}{(n+l+1)(n-l-4)} \ r^{n-4}.
\end{split}
\end{equation}
The relations, regarding the symmetric tracefree tensors (STF tensors) employed for the previous computation directly apply here again. For clarity reasons we wish to introduce a notation which will allow us to present the important intermediate results in a more concise way,
\begin{equation}
\begin{split}
I^{ab}_{II}\,=&\,\frac{63 \epsilon G}{4\pi c^2 }\sum_A\sum_{B\neq A} \tilde{m}_A\tilde{m}_B \ J^{ab}_{II},
\quad \ \ S^{ab}_{II}\,=\,\frac{63 \epsilon G}{4\pi c^2 }\sum_A\sum_{B\neq A} \tilde{m}_A\tilde{m}_B \ W^{ab}_{II}.
\end{split}
\end{equation}
In what follows we will work out the total thirty-two integrals by using the methods presented above,
\begin{equation*}
\hspace{-1.2cm}\begin{split}
J^{ab}_{II1}\,=&\,\int_{\mathcal{M}_y}  \frac{d\textbf{y}}{|\textbf{y}-\textbf{r}_{AB}|^5}  \ \frac{y^4}{y^5} \ y^a y^b\,=\,\Big[\frac{2\pi}{5} \ n^{\langle ab\rangle}+\frac{5\pi}{3}  \ \delta^{ab}\Big]r^{-1}_{AB},\\
J^{ab}_{II2}\,=&\,\int_{\mathcal{M}_y}   \frac{d\textbf{y}}{|\textbf{y}-\textbf{r}_{AB}|^5}    \ \frac{y^4}{y^5} \ y^a r_B^b\,=\,\frac{7\pi}{9} \  n^{\langle a\rangle} r^b_B \ r^{-2}_{AB},\\
J^{ab}_{II3}\,=&\,\int_{\mathcal{M}_y} \frac{d\textbf{y}}{|\textbf{y}-\textbf{r}_{Ab}|^5}  \ \frac{y^4}{y^5} \ r_B^a y^b\,=\, \frac{7\pi}{9}  \ n^{\langle b\rangle} r^a_B \ r^{-2}_{AB},\\
J^{ab}_{II4}\,=&\,\int_{\mathcal{M}_y} \frac{d\textbf{y}}{|\textbf{y}-\textbf{r}_{Ab}|^5} \ r_B^a r_B^b\,=\,\frac{10\pi}{3}  \  r_B^a r_B^b \ r^{-3}_{AB},\\
J^{ab}_{II5}\,=&\,\int_{\mathcal{M}_y} \frac{d\textbf{y}}{|\textbf{y}-\textbf{r}_{AB}|^5}   \ \frac{y^2 r_{AB}^p}{y^5} \ y^p y^a y^b\,
=\,\Big[\frac{22\pi}{105}  \ n^{\langle abp\rangle}r_{ABp}+ \frac{7\pi}{45} \ \big(\delta^{ab} n^pr_{ABp}+r^a_{AB} n^b+r^b_{AB} n^a\big)\Big] r^{-2}_{AB},\\
J^{ab}_{II6}\,=&\,\int_{\mathcal{M}_y} \frac{d\textbf{y}}{|\textbf{y}-\textbf{r}_{AB}|^5}  \ \frac{y^2 r_{AB}^p}{y^5} \ y^p r_B^a y^b\,=\,\Big[ \frac{9\pi}{25}  \ n^{\langle bp\rangle}r_{ABp}r^a_B+ \frac{10\pi}{3} \   r^b_{AB}r^a_B\Big] r^{-3}_{AB},\\
J^{ab}_{II7}\,=&\,\int_{\mathcal{M}_y} \frac{d\textbf{y}}{|\textbf{y}-\textbf{r}_{AB}|^5}  \ \frac{y^2 r_{AB}^p}{y^5} \ y^p r_B^b y^a\,=\,\Big[\frac{9\pi}{25}  \ n^{\langle ap\rangle}r_{ABp}r^b_B+ \frac{10\pi}{3} \  r^a_{AB}r^b_B\Big] r^{-3}_{AB},\\
J^{ab}_{II8}\,=&\, \int_{\mathcal{M}_y} \frac{d\textbf{y}}{|\textbf{y}-\textbf{r}_{AB}|^5}  \ \frac{y^2 r_{AB}^p}{y^5} \ y^p r_B^a r_B^b\,=\,\frac{14\pi}{15}  \ n^{\langle p\rangle}r_{ABp}r^a_Br^b_B \ r^{-4}_{AB},\\
J^{ab}_{II9}\,=&\,\int_{\mathcal{M}_y} \frac{d\textbf{y}}{|\textbf{y}-\textbf{r}_{AB}|^5} \ \frac{r_{ABm}y^my^2}{y^5} \ y^a y^b\,=\,\Big[\frac{22\pi}{105}  \ n^{\langle abm\rangle}r_{ABm}+ \frac{7\pi}{45} \  \big( \delta^{ab}n^mr_{ABm} +r^a_{AB} n^b+r^b_{AB} n^a\big)\Big] r^{-2}_{AB},\\
J^{ab}_{II10}\,=&\, \int_{\mathcal{M}_y} \frac{d\textbf{y}}{|\textbf{y}-\textbf{r}_{AB}|^5}  \ \frac{r_{ABm}y^my^2}{y^5} \ y^a r_B^b\,=\,\Big[\frac{9\pi}{25} \ n^{\langle am\rangle}r^b_B r_{ABm} +\frac{10\pi}{9} \ r^b_B r^a_{AB}\Big]r^{-3}_{AB},\\
\end{split}
\end{equation*}
\begin{equation*}
\hspace{-1.5cm}\begin{split}
J^{ab}_{II11}\,=&\,\int_{\mathcal{M}_y} \frac{d\textbf{y}}{|\textbf{y}-\textbf{r}_{AB}|^5}   \ \frac{y^2y^m r_{ABm}}{y^5} \ r_B^a y^b\,=\,\Big[\frac{9\pi}{25} \ \ n^{\langle bm\rangle}r^a_Br_{ABm}+\frac{10\pi}{9}  r^a_Br^b_{AB}\Big]r^{-3}_{AB},\\
J^{ab}_{II12}\,=&\, \int_{\mathcal{M}_y} \frac{d\textbf{y}}{|\textbf{y}-\textbf{r}_{AB}|^5} \ \frac{r_{ABm} y^my^2}{y^5} \ r_B^a r_B^b\,=\,\frac{14\pi}{15}  \  r_B^a r_B^b \ n^m r_{ABm} r^{-4}_{AB},\\
\end{split}
\end{equation*}
\begin{equation*}
\hspace{-1.6cm}\begin{split}
J^{ab}_{II13}\,=&\,\int_{\mathcal{M}_y} \frac{d\textbf{y}}{|\textbf{y}-\textbf{r}_{AB}|^5}   \ \frac{r_{ABm} r_{ABp}}{y^5} \ y^m y^p y^a y^b\,=\,r^{-3}_{AB}\Bigg[\frac{26\pi}{189}  \ n^{\langle abmp\rangle}r_{ABm}r_{ABp}+\frac{2\pi}{9} \delta^{ab}r^2_{AB}+\frac{4\pi}{9}r^a_{AB}r^b_{AB}+\\
&\,\frac{9\pi}{175} \Big[  \delta^{ab}n^{\langle mp\rangle }r_{ABm}r_{ABp}+2r^a_{AB}r_{ABp}n^{\langle bp\rangle}+2r^b_{AB}r_{ABp}n^{\langle ap\rangle}+r^2_{AB}n^{\langle ab\rangle}\Big]\Bigg],\\
\end{split}
\end{equation*}
\begin{equation*}
\hspace{-1.6cm}\begin{split}
J^{ab}_{II14}\,=&\,\int_{\mathcal{M}_y} \frac{d\textbf{y}}{|\textbf{y}-\textbf{r}_{AB}|^5}  \ \frac{r_{ABm}r_{ABp}}{y^5} \ y^m y^p r_B^a y^b\\
=&\,\Big[\frac{11\pi}{49}  \ n^{\langle mbp\rangle}r_{ABm} r_{ABp}r^a_B+\frac{14\pi}{75}\ \big(r^b_{AB} r_{ABp}r^a_Bn^p+r^2_{AB}r^a_B n^b+n^mr_{ABm} r^b_{AB}r^a_B\big)\Big] r^{-4}_{AB},\\
\end{split}
\end{equation*}
\begin{equation*}
\hspace{-1.6cm}\begin{split}
J^{ab}_{II15}\,=&\,\int_{\mathcal{M}_y} \frac{d\textbf{y}}{|\textbf{y}-\textbf{r}_{AB}|^5} \ \frac{r_{ABm}r_{ABp}}{y^5} \ y^m y^p r_B^b y^a\\
=&\,\Big[\frac{11\pi}{49} \ n^{\langle map\rangle}r_{ABm} r_{ABp}r^b_B+\frac{14\pi}{75}\big( r^a_{AB} r_{ABp}r^b_Bn^p+r^2_{AB}r^b_B\ n^a+r_{ABm} r^a_{AB}r^b_Bn^m\big)\Big]r^{-4}_{AB},\\
\end{split}
\end{equation*}
\begin{equation*}
\hspace{-1.6cm}\begin{split}
J^{ab}_{II16}\,=&\,\int_{\mathcal{M}_y} \frac{d\textbf{y}}{|\textbf{y}-\textbf{r}_{AB}|^5}  \ \frac{r_{ABm} r_{ABp}}{y^5} \ y^m y^p r_B^a r_B^b\,=\, \Big[\frac{8\pi}{21}  \ n^{\langle mp\rangle}r_{ABm}r_{ABp}r^a_Br^b_B+\frac{8\pi}{5}  \ r^2_{AB} \ r^a_Br^b_B\Big]r^{-5}_{AB}.
\end{split}
\end{equation*}
It should be noticed that, in order to avoid logarithmic divergences, the last result has been worked out by making use of $O\,:=\,\int_0^\mathcal{R} y^n \frac{r_<^l}{r_>^{l+4}}dy=\frac{-2l-4}{(n+l+1)(n-l-3)} r^{n-3}_{AB}$. These results are evaluated, in the context of binary systems, in the main part of this thesis in subsection \ref{UVQuadrupole}. After this we can look at the four surface integrals from $W_{II}^{ab}$,

\hspace{-2cm}\begin{minipage}{0.6\textwidth}
\begin{equation*}
\begin{split}
W^{ab}_{II1}\,=\,\int_{\partial\mathcal{M}}\textbf{r}d\textbf{S} \frac{y^2}{|\textbf{y}-\textbf{r}_{AB}|^5}\frac{y^2}{y^5} y^ay^b\,=&\,0,\\
W^{ab}_{II2}\,=\,\int_{\partial\mathcal{M}}\textbf{r}d\textbf{S} \frac{-r_{ABm}y^m}{|\textbf{y}-\textbf{r}_{AB}|^5}\frac{y^2}{y^5} y^ay^b\,\propto&\,\mathcal{R}^{-3},\\
W^{ab}_{II3}\,=\,\int_{\partial\mathcal{M}}\textbf{r}d\textbf{S} \frac{y^2}{|\textbf{y}-\textbf{r}_{AB}|^5}\frac{-r_{ABp}y^p}{y^5} y^ay^b\,\propto&\,\mathcal{R}^{-3},\\
W^{ab}_{II4}\,=\,\int_{\partial\mathcal{M}}\textbf{r}d\textbf{S} \frac{-r_{ABm}y^m}{|\textbf{y}-\textbf{r}_{AB}|^5}\frac{-r_{ABp}y^p}{y^5} y^ay^b\,\propto&\,\mathcal{R}^{-4},\\
\end{split}
\end{equation*}
\end{minipage}
\begin{minipage}{0.4\textwidth}
\begin{equation*}
\begin{split}
W^{ab}_{II5}\,=\,\int_{\partial\mathcal{M}}\textbf{r}d\textbf{S} \frac{y^2}{|\textbf{y}-\textbf{r}_{AB}|^5}\frac{y^2}{y^5} r_B^ay^b\,\propto&\,\mathcal{R}^{-3},\\
W^{ab}_{II6}\,=\,\int_{\partial\mathcal{M}}\textbf{r}d\textbf{S} \frac{-r_{ABm}y^m}{|\textbf{y}-\textbf{r}_{AB}|^5}\frac{y^2}{y^5} r_B^ay^b\,=&\,0,\\
W^{ab}_{II7}\,=\,\int_{\partial\mathcal{M}}\textbf{r}d\textbf{S} \frac{y^2}{|\textbf{y}-\textbf{r}_{AB}|^5}\frac{-r_{ABp}y^p}{y^5} r_B^ay^b\,=&\,0,\\
W^{ab}_{II8}\,=\,\int_{\partial\mathcal{M}}\textbf{r}d\textbf{S} \frac{-r_{ABm}y^m}{|\textbf{y}-\textbf{r}_{AB}|^5}\frac{-r_{ABp}y^p}{y^5} r_B^ay^b\,\propto&\,\mathcal{R}^{-5},\\
\end{split}
\end{equation*}
\end{minipage}

\hspace{-2cm}\begin{minipage}{0.6\textwidth}
\begin{equation*}
\begin{split}
W^{ab}_{II9}\,=\,\int_{\partial\mathcal{M}}\textbf{r}d\textbf{S} \frac{y^2}{|\textbf{y}-\textbf{r}_{AB}|^5}\frac{y^2}{y^5} y^ar_B^b\,\propto&\,\mathcal{R}^{-3},\\
W^{ab}_{II10}\,=\,\int_{\partial\mathcal{M}}\textbf{r}d\textbf{S} \frac{-r_{ABm}y^m}{|\textbf{y}-\textbf{r}_{AB}|^5}\frac{y^2}{y^5} y^ar_B^b\,=&\,0,\\
W^{ab}_{II11}\,=\,\int_{\partial\mathcal{M}}\textbf{r}d\textbf{S} \frac{y^2}{|\textbf{y}-\textbf{r}_{AB}|^5}\frac{-r_{ABp}y^p}{y^5} y^ar_B^b\,=&\,0,\\
W^{ab}_{II12}\,=\,\int_{\partial\mathcal{M}}\textbf{r}d\textbf{S} \frac{-r_{ABm}y^m}{|\textbf{y}-\textbf{r}_{AB}|^5}\frac{-r_{ABp}y^p}{y^5} y^ar_B^b\,\propto&\,\mathcal{R}^{-5},\\
\end{split}
\end{equation*}
\end{minipage}
\begin{minipage}{0.4\textwidth}
\begin{equation*}
\begin{split}
W^{ab}_{II13}\,=\,\int_{\partial\mathcal{M}}\textbf{r}d\textbf{S} \frac{y^2}{|\textbf{y}-\textbf{r}_{AB}|^5}\frac{y^2}{y^5} r_B^ar_B^b\,&\,0,\\
W^{ab}_{II14}\,=\,\int_{\partial\mathcal{M}}\textbf{r}d\textbf{S} \frac{-r_{ABm}y^m}{|\textbf{y}-\textbf{r}_{AB}|^5}\frac{y^2}{y^5} r_B^ar_B^b\,\propto&\,\mathcal{R}^{-5},\\
W^{ab}_{II15}\,=\,\int_{\partial\mathcal{M}}\textbf{r}d\textbf{S} \frac{y^2}{|\textbf{y}-\textbf{r}_{AB}|^5}\frac{-r_{ABp}y^p}{y^5} r_B^ar_B^b\,\propto&\,\mathcal{R}^{-5},\\
W^{ab}_{II16}\,=\,\int_{\partial\mathcal{M}}\textbf{r}d\textbf{S} \frac{-r_{ABm}y^m}{|\textbf{y}-\textbf{r}_{AB}|^5}\frac{-r_{ABp}y^p}{y^5} r_B^ar_B^b\,=&\,0.\\
\end{split}
\end{equation*}
\end{minipage}

$\newline$

We see that the integrals either vanish or that they are proportional to the near zone cut-off parameter $\mathcal{R}$.

\end{appendix}


\begin{thebibliography}{1}

\bibitem{Einstein1}
A. Einstein, Sitzungsber. K. Preuss. Akad. Wiss. (Berlin), 844 (1915).

\bibitem{Einstein2}
A. Einstein, Sitzungsber. K. Preuss. Akad. Wiss. 1, 688 (1916).

\bibitem{Einstein3}
A. Einstein, Sitzungsber. K. Preuss. Akad. Wiss. 1, 154 (1918).


\bibitem{Alain1}
A. Dirkes, {\it The relaxed Einstein equations in the context of a mixed UV-IR modified theory of gravity}, \href{http://iopscience.iop.org/article/10.1088/1361-6382/aa6061/meta}{Class. Quantum Grav. 34 065008 (2017)}.


\bibitem{Alain0A}
A. Dirkes, {\it Degravitation and the relaxed Einstein Equations}, \href{https://arxiv.org/abs/1604.06016}{arXiv:1604.06016 [gr-qc]}.

\bibitem{Alain0B}
A. Dirkes, {\it Degravitation, Orbital Dynamics and the Effective Barycentre}, \href{arXiv:1607.04123 [gr-qc]}{https://arxiv.org/abs/1607.04123}.




\bibitem{Taylor1}
R. A. Hulse, J. H. Taylor, {\it Discovery of a pulsar in a binary-system}, \href{http://adsabs.harvard.edu/abs/1975ApJ...195L..51H}{Astrophys. J. 195:L51-L53 (1975)}.

\bibitem{Burgay1}
M. Burgay et al., {\it An increased estimate of the merger rate of double neutron stars from observations of a highly relativistic system}, \href{http://www.nature.com/nature/journal/v426/n6966/full/nature02124.html}{Nature 426 (2003) 531-533}, \href{http://arxiv.org/abs/astro-ph/0312071}{arXiv:astro-ph/0312071}.


\bibitem{Stairs1}
I. H. Stairs, {\it Testing General Relativity with Pulsar Timing}, \href{http://relativity.livingreviews.org/Articles/lrr-2003-5/}{LivingRev.Rel.6:5, 2003}, \href{http://arxiv.org/abs/astro-ph/0307536}{arXiv:astro-ph/0307536}.

\bibitem{Stairs2}
I. H. Stairs, S. E. Thorsett, J. H. Taylor, A. Wolszczan, {\it Studies of the Relativistic Binary Pulsar PSR B1534+12: I. Timing Analysis}, \href{http://iopscience.iop.org/article/10.1086/344157/meta}{Astrophys.J. 581 (2002) 501-508}, \href{http://arxiv.org/abs/astro-ph/0208357}{arXiv:astro-ph/0208357}.

\bibitem{LIGO1}
B. P. Abbott et al. (LIGO Scientific Collaboration and Virgo Collaboration), {\it Observation of Gravitational Waves from a Binary Black Hole Merger}, \href{http://journals.aps.org/prl/abstract/10.1103/PhysRevLett.116.061102}{Phys. Rev. Lett. 116, 061102 (2016)}, \href{http://arxiv.org/abs/1602.03837}{arXiv:1602.03837 [gr-qc]}.

\bibitem{LIGO2}
B. P. Abbott et al. (LIGO Scientific Collaboration and Virgo Collaboration), {\it GW150914: The Advanced LIGO Detectors in the Era of First Discoveries}, \href{http://journals.aps.org/prl/abstract/10.1103/PhysRevLett.116.131103}{Phys. Rev. Lett. 116, 131103 (2016)}, \href{http://arxiv.org/abs/1602.03838}{ arXiv:1602.03838 [gr-qc]}.

\bibitem{LIGO3}
B. P. Abbott et al. (LIGO Scientific Collaboration and Virgo Collaboration), {\it GW150914: First results from the search for binary black hole coalescence with Advanced LIGO}, \href{http://journals.aps.org/prd/abstract/10.1103/PhysRevD.93.122003}{Phys. Rev. D 93, 122003 (2016)}, \href{http://arxiv.org/abs/1602.03839}{arXiv:1602.03839 [gr-qc]}.

\bibitem{LIGO4}
B. P. Abbott et al. (LIGO Scientific Collaboration and Virgo Collaboration), {\it GW151226: Observation of Gravitational Waves from a 22-Solar-Mass Binary Black Hole Coalescence}\href{https://journals.aps.org/prl/abstract/10.1103/PhysRevLett.116.241103}{Phys. Rev. Lett. 116, 241103 (2016)}, \href{https://arxiv.org/abs/1606.04855}{arXiv:1606.04855 [gr-qc]}.

\bibitem{LIGO5}
B. P. Abbott et al. (LIGO Scientific Collaboration and Virgo Collaboration), {\it GW170104: Observation of a 50-Solar-Mass Binary Black Hole Coalescence at Redshift 0.2}, \href{https://journals.aps.org/prl/abstract/10.1103/PhysRevLett.118.221101}{Phys. Rev. Lett., 118(22):221101 (2017)}, \href{https://arxiv.org/abs/1706.01812}{arXiv:1706.01812 [gr-qc]}.

\bibitem{LIGO6}
B. P. Abott et al. (LIGO Scientific Collaboration and Virgo Collaboration), {\it GW170814: A Three-Detector Observation of Gravitational Waves from a Binary Black Hole Coalescence}, \href{https://journals.aps.org/prl/abstract/10.1103/PhysRevLett.119.141101}{Phys. Rev. Lett. 119, 141101 (2017)}, \href{https://arxiv.org/abs/1709.09660}{arXiv:1709.09660 [gr-qc]}.

\bibitem{LIGO7}
B. P. Abott et al. (LIGO Scientific Collaboration and Virgo Collaboration), {\it GW170817: Observation of Gravitational Waves from a Binary Neutron Star Inspiral}, \href{https://journals.aps.org/prl/abstract/10.1103/PhysRevLett.119.161101}{Phys. Rev. Lett. 119, 161101}, \href{https://arxiv.org/abs/1710.05832}{
arXiv:1710.05832 [gr-qc]}.


\bibitem{Dvali1}
N. Arkani-Hamed, S. Dimopoulos, Gia Dvali, G. Gabadadze, {\it Non-Local Modification of Gravity and the Cosmological Constant Problem}, \href{http://arxiv.org/abs/hep-th/0209227v1}{arXiv:hep-th/0209227v1}.

\bibitem{Dvali2}
G. Dvali, S. Hofmann, J. Khoury, {\it Degravitation of the cosmological constant and graviton width}, \href{https://journals.aps.org/prd/abstract/10.1103/PhysRevD.76.084006}{Phys. Rev. D 76, 084006 (2007)}, \href{http://arxiv.org/abs/hep-th/0703027}{arXiv:hep-th/0703027}.

\bibitem{Barvinsky1}
A. O. Barvinsky, {\it Nonlocal action for long-distance modifications of gravity theory}, \href{http://www.sciencedirect.com/science/article/pii/S0370269303013182?via\%3Dihub}{Phys. Lett. B 572 (2003) 109-116}, \href{http://arxiv.org/abs/hep-th/0304229}{arXiv:hep-th/0304229}.

\bibitem{Barvinsky2}
A. O. Barvinsky, {\it Covariant long-distance modifications of Einstein theory and strong coupling problem}, \href{https://journals.aps.org/prd/abstract/10.1103/PhysRevD.71.084007}{Phys. Rev. D 71, 084007 (2005)}, \href{http://arxiv.org/abs/hep-th/0501093}{arXiv:hep-th/0501093}.

\bibitem{WagonerWill1976}
R. V. Wagoner, C. M. Will, {\it Post-Newtonian gravitational radiation from orbiting point masses}, \href{http://adsabs.harvard.edu/doi/10.1086/154886}{Astrophysical Journal, vol. 210, Dec. 15, 1976, pt. 1, p. 764-775}.

\bibitem{WillWiseman}
C. M. Will, Alan G. Wiseman, {\it Gravitational radiation from compact binary systems: Gravitational waveforms and energy loss to second post-Newtonian order}, \href{http://dx.doi.org/10.1103/PhysRevD.54.4813}{Phys. Rev. D 54, 4813 (1996)}, \href{http://arxiv.org/abs/gr-qc/9608012}{arXiv:gr-qc/9608012}.

\bibitem{PatiWill1}
M. E. Pati, C. M. Will, {\it PostNewtonian gravitational radiation and equations of motion via direct integration of the relaxed Einstein equations. 1. Foundations}, \href{http://dx.doi.org/10.1103/PhysRevD.62.124015}{Phys. Rev. D 62, 124015 (2000)}, \href{http://arxiv.org/abs/gr-qc/0007087}{ arXiv:gr-qc/0007087 }. 

\bibitem{PatiWill2}
M. E. Pati, C. M. Will, {\it PostNewtonian gravitational radiation and equations of motion via direct integration of the relaxed Einstein equations. 2. Two-body equations of motion to second postNewtonian order, and radiation reaction to 3.5 postNewtonia order}, \href{http://dx.doi.org/10.1103/PhysRevD.65.104008}{Phys. Rev. D 65, 104008 (2002)}, \href{http://arxiv.org/abs/gr-qc/0201001}{arXiv:gr-qc/0201001}. 

\bibitem{LandauLifshitz}
L. D. Landau, E. M. Lifshitz, The Classical Theory of Fields (Volume 2 of A Course of Theoretical Physics) Pergamon Press (1971). 

\bibitem{MisnerThroneWheeler}
C. W. Misner, K. S. Throne, J. A. Wheeler, Gravitation, Palgrave Macmillan (1973).

\bibitem{PoissonWill}
E. Poisson, C. M. Will, {\it Gravity (Newtonian, Post-Newtonian, Relativistic)}, Cambridge University Press (2014).

\bibitem{Poisson1}
E. Poisson, {The Motion of Point Particles in Curved Spacetime}, \href{https://link.springer.com/article/10.12942\%2Flrr-2004-6}{Living Rev. Relativity 7 (2004), 6 }, \href{http://arxiv.org/abs/gr-qc/0306052}{arXiv:gr-qc/0306052}.

\bibitem{Blanchet1}
L. Blanchet, {\it Gravitational Radiation from Post Newtonian Sources and Inspiralling Compact Binaries}, \href{http://relativity.livingreviews.org/Articles/lrr-2014-2/}{Living Rev. Relativity, 17, (2014), 2}, \href{https://arxiv.org/abs/1310.1528}{arXiv:1310.1528 [gr-qc]}.

\bibitem{Maggiore1}
M. Maggiore, {\it Gravitational Waves: Theory and Experiments}, Oxford University Press (2014).

\bibitem{Buonanno1}
A. Buonanno, {\it Gravitational Waves}, \href{http://arxiv.org/abs/0709.4682}{arXiv:0709.4682 [gr-qc]}.

\bibitem{Blanchet2}
L. Blanchet, G. Faye, {\it Hadamard regularization}, \href{http://dx.doi.org/10.1063/1.1308506 }{J. of Math. Phys. 41, 7675 (2000)}, \href{http://arxiv.org/abs/gr-qc/0004008}{arXiv:gr-qc/0004008}.

\bibitem{Blanchet3}
O. Poujade, L. Blanchet, {\it Post-Newtonian approximation for isolated systems calculated by matched asymptotic expansion}, \href{http://journals.aps.org/prd/abstract/10.1103/PhysRevD.65.124020}{Phys.Rev. D65 (2002) 124020}, \href{http://arxiv.org/abs/gr-qc/0112057}{arXiv:gr-qc/0112057}.

\bibitem{Blanchet4}
L. Blanchet, {\it Post-Newtonian theory and the two body problem}, \href{http://link.springer.com/chapter/10.1007\%2F978-90-481-3015-3_5}{Fundam.Theor.Phys. 162 (2011) 125-166}, \href{http://arxiv.org/abs/0907.3596}{arXiv:0907.3596 [gr-qc]}.

\bibitem{Maggiore2}
M. Jaccard, M. Maggiore, E. Mitsou, {\it Nonlocal theory of massive gravity}, \href{http://journals.aps.org/prd/abstract/10.1103/PhysRevD.88.044033}{Phys. Rev. D 88, 044033 (2013)}, \href{https://arxiv.org/abs/1305.3034}{arXiv:1305.3034 [hep-th]}.

\bibitem{Barvinsky3}
A. O. Barvinsky, G. A. Vilkovisky, {\it Beyond the Schwinger-Dewitt Technique: Converting Loops Into Trees and In-In Currents}, \href{http://www.sciencedirect.com/science/article/pii/055032138790681X}{Nucl.Phys. B282 (1987) 163-188 }.

\bibitem{Barvinsky4}
A. O. Barvinsky, G. A. Vilkovisky, {\it Covariant perturbation theory. 2: Second order in the curvature. General algorithms }, \href{http://www.sciencedirect.com/science/article/pii/055032139090047H}{Nucl.Phys. B333 (1990) 471-511 }.

\bibitem{Will1}
C. M. Will. {\it The confrontation between General Relativity and Experiment}, \href{https://link.springer.com/article/10.12942\%2Flrr-2014-4}{Living Rev. Relativity, 17, (2014), 4}, \href{http://arxiv.org/abs/1403.7377}{arXiv:1403.7377 [gr-qc]}. 

\bibitem{Esposito1}
C. Deffayet, G. Esposito-Far\`{e}se, R. P. Woodard, {\it Field equations and cosmology for a class of nonlocal metric models of MOND}, \href{https://journals.aps.org/prd/abstract/10.1103/PhysRevD.90.064038}{Phys. Rev. D 90, 089901 (2014)}, \href{http://arxiv.org/abs/1405.0393v1}{arXiv:1405.0393v1}.

\bibitem{Clifton1}
T. Clifton, P. G. Ferreira, A. Padilla , Constantinos Skordis, {\it  Modified gravity and cosmology}, \href{http://www.sciencedirect.com/science/article/pii/S0370157312000105?via\%3Dihub}{Physics Reports 513 (2012) 1-189}, \href{http://arxiv.org/abs/1106.2476}{arXiv:1106.2476}.

\bibitem{Tsujikawa1}
A. De Felice, S. Tsujikawa, {\it f($R$) Theories}, \href{https://link.springer.com/article/10.12942\%2Flrr-2010-3}{Living Rev. Rel. 13: 3, 2010}, \href{http://arxiv.org/abs/1002.4928}{arXiv:1002.4928 [gr-qc]}.

\bibitem{Woodard1}
R. P. Woodard, {\it Nonlocal Models of Cosmic Acceleration}, \href{https://link.springer.com/article/10.1007\%2Fs10701-014-9780-6}{Found Phys (2014) Volume 44, Issue2 pp213-233}, \href{https://arxiv.org/abs/1401.0254}{arXiv:1401.0254 [astro-ph.CO]}.

\bibitem{BertiBuonannoWill}
E. Berti, A. Buonanno, Clifford M. Will, {\it Testing general relativity and probing the merger history of massive black holes with LISA}, \href{http://iopscience.iop.org/article/10.1088/0264-9381/22/18/S08/meta}{Class.Quant.Grav. 22 (2005) S943-S954 }, \href{http://arxiv.org/abs/gr-qc/0504017}{arXiv:gr-qc/0504017}.

\bibitem{Modesto1}
L. Modesto, {\it Super-renormalizable quantum gravity}, \href{http://journals.aps.org/prd/abstract/10.1103/PhysRevD.86.044005}{Phys. Rev. D 86, 044005 (2012)}, \href{https://arxiv.org/abs/1107.2403}{arXiv:1107.2403 [hep-th]}.

\bibitem{Modesto2}
L. Modesto, J. W. Moffat, P. Nicolini, {\it Black holes in an ultraviolet complete quantum gravity}, \href{http://www.sciencedirect.com/science/article/pii/S0370269310013213}{Phys.Lett.B695:397-400 (2011)}, \href{https://arxiv.org/abs/1010.0680}{arXiv:1010.0680 [gr-qc]}.






\bibitem{Tomboulis1}
E. T. Tomboulis, {\it Superrenormalizable gauge and gravitational theories}, \href{https://arxiv.org/abs/hep-th/9702146}{arXiv:hep-th/9702146}.


\bibitem{Tomboulis2}
E. T. Tomboulis, {\it Renormalization and unitarity in higher derivative and nonlocal gravity theories}, \href{http://www.worldscientific.com/doi/abs/10.1142/S0217732315400052}{Mod. Phys. Lett. A, 30, 1540005 (2015)}.


\bibitem{Tomboulis3}
E. T. Tomboulis, {\it Nonlocal and quasi-local field theories}, \href{https://journals.aps.org/prd/abstract/10.1103/PhysRevD.92.125037}{Phys. Rev. D 92, 125037 (2015)}, \href{https://arxiv.org/abs/1507.00981}{arXiv:1507.00981 [hep-th]}.


\bibitem{PaisUhlenbeck1}
A. Pais, G. E. Uhlenbeck, {\it On Field Theories with Non-Localized Action}, \href{https://journals.aps.org/pr/abstract/10.1103/PhysRev.79.145}{Phys. Rev. 79, 145 (1950)}.

\bibitem{Yukawa1}
H. Yukawa, {\it Quantum Theory of Non-Local Fields. Part I. Free Fields}, \href{https://journals.aps.org/pr/abstract/10.1103/PhysRev.77.219}{Phys. Rev. 77, 219 (1950)}.


\bibitem{Modesto3}
Y.-D. Li, L. Modesto, L. Rachwal, {\it Exact solutions and spacetime singularities in nonlocal gravity}, \href{http://link.springer.com/article/10.1007\%2FJHEP12\%282015\%29173}{JHEP 12 (2015) 173}, \href{https://arxiv.org/abs/1506.08619}{arXiv:1506.08619 [hep-th]}.

\bibitem{Modesto4}
L. Modesto, L. Rachwal, {\it Universally finite gravitational and gauge theories}, \href{http://www.sciencedirect.com/science/article/pii/S0550321315003144}{Nuc. Phys. B 900 (2015) 147-169}, \href{https://arxiv.org/abs/1503.00261}{arXiv:1503.00261 [hep-th]}.

\bibitem{Modesto5}
G. Calcagni, L. Modesto, {\it Nonlocal quantum gravity and M-theory}, \href{http://journals.aps.org/prd/abstract/10.1103/PhysRevD.91.124059}{
Phys. Rev. D 91, 124059 (2015)}, \href{https://arxiv.org/abs/1404.2137}{arXiv:1404.2137 [hep-th]}.

\bibitem{Modesto6}
L. Modesto, P. Nicolini, {\it Charged rotating noncommutative black holes}, \href{https://journals.aps.org/prd/abstract/10.1103/PhysRevD.82.104035}{Phys.Rev.D82:104035 (2010)}, \href{https://arxiv.org/abs/1005.5605}{arXiv:1005.5605 [gr-qc]}.

\bibitem{Modesto7}
F. Briscese, L. Modesto, S. Tsujikawa, {\it Super-renormalizable or finite completion of the Starobinsky theory}, \href{https://journals.aps.org/prd/abstract/10.1103/PhysRevD.89.024029}{Phys.Rev. D89, (2014) 024029}, \href{}{arXiv:1308.1413 [hep-th]}.


\bibitem{Modesto8}
G. Calcagni, L. Modesto, {\it Nonlocality in string theory}, \href{http://iopscience.iop.org/article/10.1088/1751-8113/47/35/355402/meta;jsessionid=F1293A92C994B44BB2152562E1682F00.c4.iopscience.cld.iop.org}{J. Phys. A: Math. Theor. 47 (2014) 355402}, \href{https://arxiv.org/abs/1310.4957}{arXiv:1310.4957 [hep-th]}.


\bibitem{Modesto9}
A. Belenchia, D. M.T. Benincasa, A. Marciano, L. Modesto, {\it Spectral Dimension from Nonlocal Dynamics on Causal Sets}, \href{https://journals.aps.org/prd/abstract/10.1103/PhysRevD.93.044017}{Phys. Rev. D 93, 044017 (2016)}, \href{https://arxiv.org/abs/1507.00330}{arXiv:1507.00330 [gr-qc]}.

\bibitem{UVNonlocal}
S. Talaganis, T. Biswas, A. Mazumdar, {\it Towards understanding the ultraviolet behavior of quantum loops in infinite-derivative theories of gravity}, \href{http://iopscience.iop.org/article/10.1088/0264-9381/32/21/215017/meta;jsessionid=C75D80480D380BABD6C928065B729B95.c1.iopscience.cld.iop.org}{Class. Quant. Grav. 32 (2015) no. 21, 215017}, \href{https://arxiv.org/abs/1412.3467}{arXiv:1412.3467 [hep-th]}.


\bibitem{HamberWilliams1}
H. W. Hamber, R. M. Williams, {\it Nonlocal Effective Gravitational Field Equations and the Running of Newton's G}, \href{https://journals.aps.org/prd/abstract/10.1103/PhysRevD.72.044026}{Phys.Rev.D72:044026,2005}, \href{https://arxiv.org/abs/hep-th/0507017}{arXiv:hep-th/0507017}.

\bibitem{HamberWilliams2}
H. W. Hamber, R. M. Williams, {\it Constraints on Gravitational Scaling Dimensions from Non-Local Effective Field Equations}, \href{http://www.sciencedirect.com/science/article/pii/S0370269306013487?via\%3Dihub}{Phys.Lett.B643:228-234 (2006)}, \href{https://arxiv.org/abs/gr-qc/0607131}{arXiv:gr-qc/0607131}.

\bibitem{Elizalde1}
E. Elizalde, {\it Aspects of Nonlocal Models of Modified Einsteinian Gravity}, \href{http://inspirehep.net/record/1292009}{PoS ICMP2013 (2013) 002}.

\bibitem{Elizalde2}
E. Elizalde, E. O. Pozdeeva, S. Yu. Vernov, Y.-L. Zhang, {\it Cosmological Solutions of a Nonlocal Model with a Perfect Fluid}, \href{http://iopscience.iop.org/article/10.1088/1475-7516/2013/07/034/meta}{JCAP 1307 (2013) 034}, \href{https://arxiv.org/abs/1302.4330}{arXiv:1302.4330 [hep-th]}.

\bibitem{Woodard2}
S. Deser, R. P. Woodard, {\it Nonlocal Cosmology}, \href{https://journals.aps.org/prl/abstract/10.1103/PhysRevLett.99.111301}{Phys. Rev. Lett. 99 111301 (2007)}, \href{https://arxiv.org/abs/0706.2151}{arXiv:0706.2151 [astro-ph]}.

\bibitem{Odintsov}
Shin'ichi Nojiri, Sergei D. Odintsov, {\it Unified cosmic history in modified gravity: from F(R) theory to Lorentz non-invariant models}, \href{Phys. Rept.505:59-144,2011}{Phys.Rept.505:59-144, 2011}, \href{https://arxiv.org/abs/1011.0544v4}{arXiv:1011.0544 [gr-qc]}.

\bibitem{Esposito2}
C. Deffayet, G. Esposito-Farese, R. P. Woodard, {\it Nonlocal metric formulations of MOND with sufficient lensing}, \href{https://journals.aps.org/prd/abstract/10.1103/PhysRevD.84.124054}{Phys. Rev. D 84 124054 (2011)}, \href{https://arxiv.org/abs/1106.4984}{arXiv:1106.4984 [gr-qc]}.

\bibitem{Maggiore3}
M. Maggiore, M. Mancarella, {\it Non-local gravity and dark energy}, \href{https://journals.aps.org/prd/abstract/10.1103/PhysRevD.90.023005}{Phys. Rev. D 90, 023005 (2014)}, \href{https://arxiv.org/abs/1402.0448}{arXiv:1402.0448 [hep-th]}.


\bibitem{Maggiore4}
G. Cusin, S. Foffa, M. Maggiore, M. Mancarella, {\it Imprint of primordial inflation on the dark energy equation of state in non-local gravity}, \href{https://arxiv.org/abs/1610.05664}{arXiv:1610.05664 [hep-th]}.


\bibitem{PeskinSchroeder}
M. E. Peskin, D. V. Schroeder, {\it An introduction to Quantum Field Theory} Westview Press (1995).

\bibitem{Snyder1}
H. S. Snyder, {\it Quantized Space-Time}, \href{https://journals.aps.org/pr/abstract/10.1103/PhysRev.71.38}{Phys. Rev., 71, 38-41 (1947)}.

\bibitem{Moffat1}
J. W. Moffat, {\it Noncommutative Quantum Gravity}, \href{http://www.sciencedirect.com/science/article/pii/S0370269300010649?via\%3Dihub}{Phys.Lett. B491, 345-352 (2000)}, \href{https://arxiv.org/abs/hep-th/0007181}{arXiv:hep-th/0007181}.

\bibitem{Moffat2}
J. W. Moffat, {\it Ultraviolet Complete Quantum Gravity}, \href{https://link.springer.com/article/10.1140\%2Fepjp\%2Fi2011-11043-7}{Eur.Phys.J.Plus 126:43 (2011)}, \href{https://arxiv.org/abs/1008.2482}{arXiv:1008.2482 [gr-qc]}.

\bibitem{NoncommutativeTheta1}
A. Smailagic, E. Spallucci, {\it Feynman Path Integral on the Noncommutative Plane}, \href{http://iopscience.iop.org/article/10.1088/0305-4470/36/33/101/meta;jsessionid=A02B8C1D115A19884D7F2363972B6E2E.c1.iopscience.cld.iop.org}{J.Phys.A36:L467 (2003)}, \href{https://arxiv.org/abs/hep-th/0307217}{arXiv:hep-th/0307217}.

\bibitem{NoncommutativeTheta2}
A. Smailagic, E. Spallucci, {\it UV divergence-free QFT on noncommutative plane}, \href{http://iopscience.iop.org/article/10.1088/0305-4470/36/39/103/meta}{J.Phys. A36 (2003) L517-L521}, \href{https://arxiv.org/abs/hep-th/0308193}{arXiv:hep-th/0308193}.

\bibitem{NoncommutativeTheta3}
A. Smailagic, E. Spallucci, {\it Lorentz invariance and unitarity in UV finite NCQFT}, \href{http://iopscience.iop.org/article/10.1088/0305-4470/37/28/008/meta}{J.Phys. A37 (2004) 1-10}, \href{https://arxiv.org/abs/hep-th/0406174}{arXiv:hep-th/0406174}.

\bibitem{NoncommutativeTheta4}
P. Nicolini, A. Smailagic, E. Spallucci, {\it Noncommutative geometry inspired Schwarzschild black hole}, \href{http://www.sciencedirect.com/science/article/pii/S0370269305016126?via\%3Dihub}{Phys.Lett.B632:547-551 (2006)}, \href{https://arxiv.org/abs/gr-qc/0510112}{arXiv:gr-qc/0510112}.

\bibitem{NoncommutativeTheta5}
S. Ansoldi, P. Nicolini, A. Smailagic, E. Spallucci, {\it Noncommutative geometry inspired charged black holes}, \href{http://www.sciencedirect.com/science/article/pii/S0370269306015607?via\%3Dihub}{Phys.Lett.B645:261-266 (2007)}, \href{https://arxiv.org/abs/gr-qc/0612035}{arXiv:gr-qc/0612035}.

\bibitem{NoncommutativeTheta6}
M. Sprenger, P. Nicolini, M. Bleicher, {\it Quantum Gravity signals in neutrino oscillations}, \href{http://www.worldscientific.com/doi/abs/10.1142/S0218301311040517}{Int.J.Mod.Phys.E20:supp02,1-6 (2011)}, \href{https://arxiv.org/abs/1111.2341}{arXiv:1111.2341 [hep-ph]}.

\bibitem{Wataghin}
G. Wataghin, {\it Bemerkung über die Selbstenergie der Elektronen}, \href{https://link.springer.com/article/10.1007\%2FBF01352311}{Zeitschrift für Physik 88, 92 (1934)}.

\bibitem{Efimov1}
G. V. Efimov, {\it Non-Local Quantum Theory of the Scalar Field}, \href{https://link.springer.com/article/10.1007/BF01646357}{Commun.Math. Phys. (1967) 5: 42}


\bibitem{Efimov2}
G. V. Efimov, {\it On a Class of Relativistic Invariant Distributions}, \href{https://link.springer.com/article/10.1007/BF01648331}{Commun.Math. Phys. (1968) 7: 138}.


\bibitem{Namsrai}
K. H. Namsrai, {\it Nonlocal Quantum Field Theory and Stochastic Quantum Mechanics}, \href{http://www.springer.com/de/book/9789027720016}{Springer-Verlag}, \href{https://books.google.lu/books?id=5Y7nCAAAQBAJ&pg=PR14&lpg=PR14&dq=non+local+quantum+field+theory+and+stochastic+quantum+mechanics&source=bl&ots=hkATuXIV49&sig=LP1pHyxb-2Ed-Fh6swAnDGqwtZw&hl=de&sa=X&ved=0ahUKEwjL7rWy37DUAhXkbZoKHenVBu4Q6AEIODAD#v=onepage&q=non\%20local\%20quantum\%20field\%20theory\%20and\%20stochastic\%20quantum\%20mechanics&f=false}{Google-books}.

\bibitem{NLSFT1}
I. Ya. Aref'eva, A. S. Koshelev, {\it Cosmic acceleration and crossing of w=-1 barrier in non-local Cubic Superstring Field Theory model}, \href{http://iopscience.iop.org/article/10.1088/1126-6708/2007/02/041/meta}{JHEP 0702:041 (2007)}, \href{https://arxiv.org/abs/hep-th/0605085}{arXiv:hep-th/0605085}.

\bibitem{NLSFT2}
A. S. Koshelev, {\it Non-local SFT Tachyon and Cosmology}, \href{http://iopscience.iop.org/article/10.1088/1126-6708/2007/04/029/meta;jsessionid=4373A174081EB0DC6E6A07AD86B9978D.c2.iopscience.cld.iop.org}{JHEP 0704:029 (2007)}, \href{https://arxiv.org/abs/hep-th/0701103}{arXiv:hep-th/0701103}.


\bibitem{Parker1}
L. Parker, A. Raval, {\it Nonperturbative effects of vacuum energy on the recent expansion of the universe}, \href{https://journals.aps.org/prd/abstract/10.1103/PhysRevD.60.063512}{Phys.Rev.D60:063512 (1999)}, \href{https://arxiv.org/abs/gr-qc/9905031}{arXiv:gr-qc/9905031}.

\bibitem{Parker2}
L. Parker, D. A. T. Vanzella, {\it Acceleration of the universe, vacuum metamorphosis, and the large-time asymptotic form of the heat kernel}, \href{https://journals.aps.org/prd/abstract/10.1103/PhysRevD.69.104009}{Phys.Rev. D69 (2004) 104009}, \href{https://arxiv.org/abs/gr-qc/0312108}{arXiv:gr-qc/0312108}.

\bibitem{TullyFisher1}
R. B. Tully, J. R. Fisher, {\it A new method of determining distances to galaxies}, \href{http://cdsads.u-strasbg.fr/cgi-bin/bib_query?1977A\%26A....54..661T}{Astronomy and Astrophysics, vol. 54, no. 3, Feb. 1977, p. 661-673}.



\bibitem{DirkesMaziashviliSilagadze}
A. R. P. Dirkes, M. Maziashvili, Z. K. Silagadze, {\it Black hole remnants due to Planck-length deformed QFT}, \href{http://www.worldscientific.com/doi/abs/10.1142/S0218271816500152}{Int. J. Mod. Phys. D25, 1650015 (2016)}, \href{https://arxiv.org/abs/1309.7427}{arXiv:1309.7427 [gr-qc]}.


\bibitem{Maziashvili2}
M. Maziashvili, {\it Field propagation in a stochastic background space: The rate of light incoherence in stellar interferometry}, \href{https://journals.aps.org/prd/abstract/10.1103/PhysRevD.86.104066}{Phys. Rev. D 86 (2012) 104066}, \href{https://arxiv.org/abs/1206.4388}{arXiv:1206.4388 [gr-qc]}.


\bibitem{Maziashvili3}
M. Maziashvili, L. Megrelidze, {\it Minimum-length deformed QM/QFT, issues and problems}, \href{https://academic.oup.com/ptep/article-lookup/doi/10.1093/ptep/ptt107}{Prog. Theor. Exp. Phys. (2013) 123B06}, \href{https://arxiv.org/abs/1212.0958}{arXiv:1212.0958 [hep-th]}.

\bibitem{Maziashvili4}
M. Maziashvili, {\it Extra-dimensional generalization of minimum-length deformed QM/QFT and some of its phenomenological consequences}, \href{https://journals.aps.org/prd/abstract/10.1103/PhysRevD.91.064040}{Phys. Rev. D 91, 064040 (2015)}, \href{https://arxiv.org/abs/1502.07535}{arXiv:1502.07535 [hep-th]}.

\bibitem{Maziashvili5}
M. Maziashvili, {\it Light incoherence due to background space fluctuations}, \href{https://journals.aps.org/prd/abstract/10.1103/PhysRevD.94.124044}{Phys. Rev. D 94, 124044 (2016)}, \href{https://arxiv.org/abs/1612.03259}{arXiv:1612.03259 [gr-qc]}.

\bibitem{Wataghin2}
G. Wataghin, {\it Quantum Theory and Relativity}, \href{http://www.nature.com/nature/journal/v142/n3591/pdf/142393b0.pdf}{Nature 142, 393-394 (1938)}.

\end{thebibliography}
\end{document}